\documentclass[useAMS, usenatbib]{mnras}
\usepackage{graphicx,amsmath,color,amssymb}

\usepackage[pdftitle={A hybrid method for simulating non-linear neutrino structure}]{hyperref}

\newcommand{\beq}{\begin{equation}}
\newcommand{\eeq}{\end{equation}}
\newcommand{\barr}{\begin{eqnarray}}
\newcommand{\earr}{\end{eqnarray}}

\newcommand{\rme}{\textrm{e}}

\newcommand{\bs}{\boldsymbol}

\renewcommand{\jcap}{JCAP}
\newcommand{\gadget}{{\small GADGET\,}}

\begin{document}

\title{An Efficient and Accurate Hybrid Method for Simulating Non-Linear Neutrino Structure}
\author[ S. Bird et al.]{Simeon Bird$^1$\thanks{E-mail: sbird@ucr.edu}, Yacine Ali-Ha\"{\i}moud$^2$, Yu Feng$^3$, Jia Liu$^4$\vspace{1.5mm}\\
$^1$University of California Riverside, Riverside, CA  and Johns Hopkins University, Baltimore, MD\\
$^2$Center for Cosmology and Particle Physics, Department of Physics,
New York University, New York, NY\\
$^3$University of California, Berkeley, CA\\
$^4$Department of Astrophysical Sciences, Princeton University, Princeton, NJ
}

\date{\today}

\pagerange{\pageref{firstpage}--\pageref{lastpage}} \pubyear{2012}
\pagenumbering{arabic}
\label{firstpage}

\maketitle

\begin{abstract}
We present an efficient and accurate method for simulating massive neutrinos in cosmological structure formation simulations, together with an easy to use public implementation. Our method builds on our earlier implementation of the linear response approximation (LRA)
for neutrinos, coupled with an $N$-body code for cold dark matter particles. The LRA's good behaviour at early times and in the linear regime is preserved, while better following the non-linear clustering of neutrinos on small scales. Massive neutrinos are split into initially ``fast'' and ``slow'' components. The fast component is followed analytically with the LRA all the way to redshift zero. The slow component is evolved with the LRA only down to a switch-on redshift $z_\nu = 1$, below which it is followed with the particle method, in order to fully account for its non-linear evolution. The slow neutrino particles are initialized at $z = 99$ in order to have accurate positions and velocities at the switch-on time, but are not used to compute the potential until $z \leq 1$, thus avoiding the worst effect of particle shot noise. We show that our hybrid method matches (and for small neutrino masses, exceeds) the accuracy of neutrino particle simulations with substantially lower particle load requirements.
\end{abstract}


\section{Introduction}

Neutrinos are the lightest standard-model fermions, and neutrino oscillation experiments have measured the differences of their masses squared to high precision \citep{Becker-Szendy_1992, Fukuda_1998, deSalas_17}. However, measuring the absolute neutrino masses in the laboratory is challenging due to the large difference in mass scales between neutrinos and other standard model particles \cite[although see][]{Wolf_2010}.


Fortunately, massive neutrinos 
also affect the growth of large-scale structure, which is mostly sensitive to the fraction of non-relativistic matter in neutrinos, hence to the sum of the neutrino masses $M_\nu$. They behave as light thermal relics, suppressing clustering below their thermal
free-streaming length \citep[e.g.][]{Lesgourgues_2006, Wong_2011}.
Measurements of the clustering of matter and matter tracers in the Universe can detect this effect and thus constrain, and eventually measure, the total mass of neutrinos.

Cosmological constraints on the neutrino mass sum $M_\nu$ are quickly approaching the lower limit implied by neutrino oscillation data, which is $M_\nu > 0.06$ eV assuming a normal hierarchy. For example, the Planck team obtained a 95 \% CL upper limit of $M_\nu<0.23$~eV~\citep{planck2015xiii} using primary cosmic microwave background (CMB) temperature data, combined with low-$\ell$ polarization, CMB lensing, type Ia supernovae~\citep{Betoule_2014}, and baryon acoustic oscillation
measurements~\citep{Beutler_2011, Anderson_2014, Ross_2015}. \cite{Palanque_2015} found a tighter constraint of $M_\nu<0.15$~eV by adding Lyman-$\alpha$ forest data from the Sloan Digital Sky Survey (SDSS), and \cite{Vagnozzi:2017} combined multiple probes to obtain a similar limit. However, recent weak-lensing data from the Dark Energy Survey combined with Planck weakens the upper limit to $0.29$ eV \citep{DES_2017}, and the most recent galaxy power spectrum measurements from SDSS show a slight preference for a non-zero neutrino mass of $M_\nu = 0.3$ eV \citep{Beutler_2014}. Interestingly, \cite{Poulin:2018} found that a preference for a neutrino mass $\sim 0.4$ eV exists in the combination of CMB, galaxy and BAO data, as long as neutrinos are constrained at the same time as an extended dark energy model.
The Dark Energy Spectroscopic Instrument (DESI) \citep{DESI} will be able to measure the neutrino mass with a $1-\sigma$ error of $0.024$ eV \citep{FontRibera:2014}. The Large Synoptic Survey Telescope~(LSST) \citep{LSST, Joudaki_2012} will constrain the neutrino mass with a comparable $1-\sigma$ error of $0.041$ eV \citep{Banerjee:2018}. The EUCLID satellite is expected to achieve a precision of $0.015$ eV \citep{Basse:2014}. Finally, next-generation CMB stage 4 experiments are expected to measure the neutrino mass via CMB lensing at a $1-\sigma$ error of $0.073$, better if the Planck measurement of the optical depth to reionization can improved upon\citep{Abazajian_16, Archidiacono:2017}.

Realizing the statistical power of future surveys will require extremely accurate modelling of structure growth and the effects of massive neutrinos on the matter density field.
Furthermore, current and future experiments achieve their statistical power from small scales where structure formation is in the non-linear regime \citep[e.g.~][]{Troxel_2017, HSC_2017}.
As following structure growth on non-linear scales ultimately requires fitting to $N$-body cosmological simulations, there is an urgent need to incorporate massive neutrinos into cosmological structure simulations
in a way both accurate and computationally inexpensive.
If the simulation methods used are inaccurate, experiments will measure incorrect values for the neutrino mass.
Conversely, if simulation techniques are overly computationally intensive, the number of simulations which can be performed will be reduced, again impeding the accuracy of the cosmological parameter measurement.

The standard approach to include massive neutrinos in structure-formation simulation has been to use the $N$-body method, which consists of solving for the evolution of discrete phase-space chunks or ``particles", akin to those used for CDM, but with a
large thermal velocity imposed in the initial conditions. This approach has been used extensively in the past \cite[e.g.~][]{Brandbyge_2008, Bird_2012, Inman_2017, FVN_2017}. Its main advantages are that it is simple to implement and fully includes non-linear clustering of neutrinos\footnote{Although note that the small-scale tree force for the neutrino particles is often disabled in previous works.}. It has been used to examine voids \citep{Massara_2015}, clusters and halos \citep{FVN_2014, Castorina_2014, Costanzi_2013}, large-scale clustering \citep{Castorina_2015} and the ISW effect \citep{Carbone_2016}.

Particle neutrinos are computationally expensive and can suffer from a variety of numerical problems related to their initially large thermal velocities. For instance, fast neutrino particles frequently move between processors in a parallel code, limiting scalability. For $M_\nu < 0.1$~eV, it is important to model the neutrino hierarchy. With particle neutrinos this requires at least two separate particle species, hence more particles for the same resolution.

Finally, in the standard method of initializing particle simulations, the initial velocity of each neutrino particle is random in magnitude and direction. After a few timesteps the positions of neutrino particles are also effectively randomized. The power spectrum of neutrino particles, $P_\nu$, will include a scale invariant white noise term, $P_\mathrm{N} \propto 1/N_\mathrm{part}$, called shot noise. At early times and on small scales this shot noise dominates over the small intrinsic perturbations of the neutrino fluid. 
The effect of shot noise on the gravitational potential may lead to numerical errors, including in some cases the formation of spurious structures \citep{Wang_White_07}.

A variety of techniques have been proposed to reduce the effects of neutrino particle shot noise.
The most obvious is to massively increase the number of neutrino particles, to which the shot noise is inversely proportional; for instance, \cite{Emberson_17} simulated $\approx 3\times 10^{12}$ particles.
\cite{Viel_2010} disabled the short-range tree force for their neutrino particles, effectively smoothing the gravitational force from neutrinos on scales of order the mean inter-particle spacing. \cite{Hannestad_2012} closed the Boltzmann hierarchy at second order, using smoothed particle hydrodynamics to model neutrinos as an approximate fluid. \cite{Banerjee_2016} used a similar method, but further reduced shot noise by softening the gravitational force on scales corresponding to the thermal free-streaming length. \cite{Dakin:2017} modelled neutrinos using a combination of linear theory and a second-order expansion of the non-linear collisionless Boltzmann equation. This is computationally efficient, but their current implementation has limited dynamic range. Finally, \cite{Banerjee_2018} alter the standard way of initializing neutrino particles. Instead of assigning each neutrino particle on a grid a random thermal velocity, they assign a predetermined specific magnitude and direction to multiple neutrino particles distributed on a coarser grid. Multiple grids are hence used to sample the Fermi-Dirac distribution, with the direction of the neutrino particles' velocity distributed as isotropically as possible at each point. This substantially suppresses shot noise, although \cite{Brandbyge:2018} showed that it can induce spurious correlations between neutrinos, especially those with initially large momenta.


While these techniques are successful in reducing shot noise, they always involve a significant additional computational cost or reduced accuracy. An orthogonal route that has been explored is to rely on perturbation theory to model massive neutrinos. For instance, \cite{Brandbyge_2009} include neutrinos via the linear theory neutrino power spectrum, and show that this is accurate to $< 1\%$ for $k < 1$ and $M_\nu < 0.3$ eV. In the same spirit, \cite{AHB}, hereafter AHB13, proposed modelling neutrinos using the linear response approximation \citep{Bond_1980, Ma_1994}. This is more accurate than pure linear theory, as it accounts for the full non-linear CDM potential to which neutrinos respond, and correctly models the phases of the neutrino overdensity, at no additional computational cost.
AHB13 showed that the linear response approximation (hereafter, LRA) accurately describes the effect of massive neutrinos on the growth of cold dark matter (CDM), computing the matter power spectrum $P(k)$ at $ < 0.2\%$ accuracy for $M_\nu < 0.6$ eV. Due to its minimal computational overhead relative to a pure CDM simulation, it has been used to investigate the combined effects of massive neutrinos and baryons \citep{Mummery_2017} and to constrain the neutrino mass using hydrodynamic simulations of large-scale structure \citep{McCarthy_2018, McCarthy_2017}.

The implementation of the LRA in AHB13 is applied to \emph{all} neutrinos, regardless of their initial thermal velocity. AHB13 found that, for $M_\nu \gtrsim 0.3$ eV and $z \lesssim 0.5$, this method under-estimates the clustering of neutrinos relative to that found in particle simulations, even though their characteristic overdensity is significantly less than unity on all scales. This can be understood qualitatively as follows: the LRA is formally the first order of an expansion in $\phi/v^2$, and is increasingly inaccurate for particles with low initial velocity $v$. As the initial velocity of neutrinos follows a Fermi-Dirac distribution, there are always neutrinos that are slow enough to cluster non-linearly. These neutrinos can dominate the clustering on small scales, even if they are a small fraction of the neutrino matter density, because linear perturbations of the bulk of neutrinos are highly suppressed.

In this paper, we present an improved hybrid method, allowing us to accurately model not only the total matter clustering but also the clustering of neutrinos.
As in AHB13, we model all neutrinos with the LRA, but only down to a cutoff redshift $z_\nu \approx 1$. Later on, neutrinos are split by their initial velocity into a fast and a slow components. The LRA is applied only to the fast component, while the slow component is followed using neutrino particles. The slow initial velocity of our particle neutrinos mitigates the numerical problems that plague purely particle simulations. Our new method allows a single simulation code to produce a well-converged neutrino simulation, at any neutrino mass, which includes late-time non-linear growth in the neutrino sector.

Our hybrid method is conceptually similar to the pioneering hybrid method of \cite{Brandbyge_2010}, hereafter BH10. In both methods fast neutrinos are followed using analytic approximations, while particles are used for sufficiently slow neutrinos below a cutoff redshift. However, our method improves on BH10 in several ways. First, we use the linear-response approximation for fast neutrinos, instead of full linear theory as in BH10. At no additional cost, this allows us to appropriately capture (i) the increased neutrino perturbations due to the deepening of gravitational potentials beyond linear theory and (ii) better approximate the phases of neutrino perturbations, which, like the CDM, de-correlate from their initial distribution due to large-scale bulk motions \citep[see e.g.][]{Tassev:2012}. Secondly, we initialize all particle neutrinos at the onset of the simulations, although we only use particle neutrinos as tracer (non-gravitating) particles until $z_\nu$. This differs from BH10, who initialize neutrinos dynamically as the thermal velocity of each neutrino bin drops below the gravitational flow velocity measured from the CDM.
By the late time at which particle neutrinos are actually used to compute density perturbations, their positions and velocities are very accurate. In particular, this allows us to capture the non-trivial correlations between thermal and bulk velocities, as opposed to BH10’s assumption that they are completely uncorrelated. Such correlations are expected because neutrinos with different thermal motions are deflected differently by gravitational potentials, even if their initial thermal velocities have the same magnitude.
Last but not least, we split neutrinos into ``slow” and “fast” particles according to their unperturbed velocity, rather than momentum. Our splitting criterion, which we justify in detail, can thus be employed, unaltered, for any neutrino mass.

We provide an improved public implementation of our neutrino simulation
method\footnote{The latest version may be found here: \url{https://github.com/sbird/kspace-neutrinos/}}.
While the implementation in AHB13 was tied to \gadget \citep{Springel_2005}, our new version is adaptable to a variety of structure simulation codes. We also include patches for \gadget-2 \citep{Springel_2005}, which were used for a large suite of simulations \cite[``MassiveNuS'': ][]{Liu_2017}\footnote{The ``MassiveNuS'' simulation data is publicly available at \url{http://columbialensing.org}}.

This article is organized as follows. In Section \ref{sec:lin_resp}, we review the linear response approximation, and study its regime of validity in detail. We describe our hybrid method to simulate neutrinos in Section \ref{sec:hybrid} and our suite of simulations in Section \ref{sec:simulations}. We describe our results and compare different methods in Section \ref{sec:results}. We conclude in Section \ref{sec:conclusion}. Appendix \ref{sec:manual} is a user's manual for our neutrino module. In appendix \ref{sec:initcond} we describe improvements to our initial conditions since AHB13.

\section{Linear-response approximation} \label{sec:lin_resp}

The linear response approximation (herafter, LRA), sometimes referred to as the \cite{Gilbert_1966} approximation, consists of linearizing the collisionless Boltzmann equation in the gravitational potential. Formally, the LRA is the first order of an expansion in $\phi/v^2$, where $v^2$ is the variance of the velocity distribution. This implies that it is only well defined for \emph{hot} species, and is not adapted to standard cold dark matter \citep{YAH_15}. It is equivalent to assuming nearly-straight-line trajectories, with small deflections of order $\phi/v^2$. The LRA was first applied to cosmological neutrinos in the linear regime by \cite{Bond_Szalay_1983} and \cite{Brandenberger_1987}, and by several authors since then \citep{Singh_Ma_2003, Ringwald_Wong_2004}. In AHB13, the LRA was integrated into an N-body code and applied to the entire neutrino distribution. As the linear response in this case is to the Newtonian potential sourced by the non-linearly clustered matter, it is possible to obtain the neutrino perturbation even in the non-linear regime of structure formation. In this section we review this approximation and study its regime of validity.

\subsection{General equations}

We start by briefly reviewing the LRA for a general collisionless species with phase-space density $f$. We work in the deep sub-horizon limit, in the conformal Newtonian gauge \citep{Ma_1995}. We denote by $s$ the ``superconformal'' time $(ds \equiv dt/a^2$, where $a$ is the scale factor), and overdots denote derivatives with respect to $s$. Comoving scales are denoted by $\bs{x}$ and $\bs{u} \equiv \dot{\bs{x}}$ denotes the rescaled peculiar velocity of a massive particle. We normalize the phase-space density such that the overdensity is given by
\beq
1 + \delta(s, \bs{x}) = \int d^3 u ~ f(s, \bs{x}, \bs{u}).
\eeq
For a non-relativistic particle, the geodesic equation is $\dot{\bs{u}} = - a^2  \bs{\nabla}_{\bs{x}} \phi$, where $\phi$ is the Newtonian potential. The phase-space density is conserved along trajectories, as is encoded by the collisionless Boltzmann (or Vlasov) equation
\beq
\dot{f} + \bs{u} \cdot \bs{\nabla}_{\bs{x}} f - a^2 \bs{\nabla}_{\bs{x}} \phi \cdot \bs{\nabla}_{\bs{u}} f = 0. \label{eq:Vlasov}
\eeq
The linear response method consists of solving for $f$ to linear order in the gravitational potential. Specifically, it is the first order in an expansion in the small parameter $\epsilon \sim a^2 \phi/u^2$. We denote by $f^0(\bs{u})$ the unperturbed, homogenous (but not necessarily isotropic) phase-space density, which integrates to unity. Linearizing Eq.~\eqref{eq:Vlasov} and taking its Fourier transform, we get, denoting by $\bs{k}$ the comoving wavenumber,
\beq
\dot{f} + i (\bs{k} \cdot \bs{u}) f = i a^2 \phi ~ \bs{k} \cdot \bs{\nabla}_{\bs{u}} f^0. \label{eq:Boltz-Fourier}
\eeq
Given initial conditions at $s_i$, this has an explicit integral solution (see AHB13),
\barr
f(s, \bs{k}, \bs{u}) &=& \rme^{- i(\bs{k} \cdot \bs{u}) (s - s_i)} f(s_i, \bs{k}, \bs{u}) \nonumber\\
&+& i \bs{k} \cdot \bs{\nabla}_{\bs{u}} f^0 \int_{s_i}^s d s' \rme^{- i \bs{k} \cdot \bs{u} (s - s')} a'^2 \phi(s', \bs{k}).~~~
\earr
The overdensity is then obtained by integrating over velocities. Using Poisson's equation, $k^2 \phi = - \frac32 H_0^2 \Omega_M a^{-1} \delta_M$, where $\delta_M$ is the total matter overdensity, we arrive at
\barr
\delta(s, \bs{k}) &=& \delta_{\rm I}(s,s_i, \bs{k})\nonumber\\
&+& \frac32 H_0^2 \Omega_M \int_{s_i}^s d s' (s-s') \mathcal{I}[\bs{k}(s-s')] a' \delta_M(s', \bs{k})\,. ~~~~\label{eq:delta-phi}
\earr
The first piece corresponds to the propagation of initial perturbations from $s_i$ to $s$,
which is much smaller than the propagation term for our massive neutrino simulations.
The kernel $\mathcal{I}$ is the Fourier transform of the unperturbed phase-space density \citep{Brandenberger_1987, Bertschinger_Watts_1988}:
\barr
\mathcal{I}(\bs{k}\Delta s ) 
= \int d^3 u ~ f^0(\bs{u})~\rme^{- i \bs{u} \cdot \bs{k} \Delta s}. \label{eq:I(k)}
\earr


\subsection{Regime of validity} \label{sec:validity}

In AHB13, the linear response approximation was used for the \emph{entire} phase-space of neutrinos. Yet, this approximation should eventually break down below some critical velocity $v_{\rm crit}$, as the behaviour of neutrinos approaches that of the CDM. In this section we estimate $v_{\rm crit}$. Whenever required, we use the non-linear matter power spectrum estimated by \textsc{class} \citep{Lesgourgues_11, Blas_11} with the \textsc{halofit} approximation \citep{Smith_2003}.

\subsubsection{Single velocity bin} \label{sec:single-bin}

We will expand the neutrino velocity distribution into bins of equal velocity, and show for which bins the LRA is valid. Let us consider a non-isotropic unperturbed phase-space distribution centered at a velocity bin $\bs{u}_0$:
\beq
f^0_{\bs{u}_0}(\bs{u}) \equiv \delta_{\rm D}(\bs{u}- \bs{u}_0),
\eeq
where $\delta_{\rm D}$ is the Dirac function\footnote{This is equivalent to the multi-fluid approach of \cite{Dupuy_14}.}. Inserting this distribution into Eqs.~\eqref{eq:delta-phi}--\eqref{eq:I(k)} and neglecting the initial condition piece, we obtain the overdensity
\barr
\delta_{\bs{u}_0}(s, \bs{k}) = \frac32 H_0^2 \Omega_M \int_{s_i}^s ds' (s - s') a' \rme^{- i \bs{u}_0 \cdot \bs{k} (s - s')} \delta_{\rm M}(s', \bs{k}).
\earr
The dimensionless power spectrum\footnote{The dimensionless power spectrum $\Delta^2(k)$ is related to the usual power spectrum $P(k)$ through $\Delta^2(k) = k^3 P(k)/(2 \pi^2)$.} of $\delta_{\bs{u}_0}$ is then anisotropic, and given by
\barr
\Delta^2_{\bs{u}_0}(s, \bs{k}) =\left(\frac32 H_0^2 \Omega_M\right)^2 \iint_{s_i}^s ds' ds'' (s - s') a' (s - s'') a'' \nonumber\\
\rme^{i \bs{u}_0 \cdot \bs{k} (s'' - s')} \Delta^2_M(s', s'', k), \label{eq:P_u0}
\earr
where $\Delta^2_M(s', s'', k)$ is the unequal-time dimensionless matter power spectrum, defined as
\barr
\langle \delta_M^*(s',\bs{k}')\delta_M(s'',\bs{k})\rangle \equiv \frac{(2 \pi)^6}{4 \pi k^3} \Delta^2_M(s', s'', k) \delta_{\rm D}(\bs{k}' - \bs{k}).
\earr
We can equivalently rewrite it in terms of the unequal-time correlation coefficient $c_M$, defined as
\beq
\Delta^2_M(s', s'', k) \equiv c_M(s', s'', k) \Delta_M(s', k) \Delta_M(s'', k),
\eeq
and such that $|c_M(s, s', k)| \leq 1$, and $c_M(s, s, k) = 1$. The variance per $k$-interval is obtained by averaging $\Delta^2_{\bs{u}_0}(\bs{k})$ over the orientations of $\bs{k}$, which is equivalent to averaging over the orientations of $\bs{u}_0$:
\barr
&&\langle \Delta^2_{\bs{u}_0}(s, \bs{k})\rangle_{\hat{k}} = \langle \Delta^2_{\bs{u}_0}(s, \bs{k})\rangle_{\hat{u}_0} = \left(\frac32 H_0^2 \Omega_M\right)^2  \nonumber\\
&&\times \iint_{s_i}^s ds' ds'' (s - s') a' (s - s'') a'' c_M(s', s'', k) \nonumber\\
&&~~~~~~~~~ j_0[u_0 k (s'' - s')]~  \Delta_M (s', k) \Delta_M(s'', k), \label{eq:P_u0-av}
\earr
where $j_0$ is the zeroth-order spherical Bessel function.

There is no simple approximation for the unequal-time correlation coefficient, which makes this integral difficult to evaluate in full generality. To simplify it, let us first note that on linear scales, $c_M \approx 1$. Secondly, on scales smaller than the free-streaming scale, i.e. $k u_0 (ds/d\ln a) \gg 1$, the spherical Bessel function oscillates on a timescale much shorter than the rest of the integrand, and may be approximated as $j_0[u_0 k (s'' - s')] \approx \frac{\pi}{u_0 k} \delta_{\rm D}(s'' - s')$. This selects mostly the equal-time correlation coefficient, equal to unity whether the matter field is linear or not. It also implies that the power per $k$ interval is proportional to $u_0^{-1}$ in this regime.
Explicitly, the non-linear scale is approximately
 \beq
 k_{\rm nl}(z) \approx 0.2~50^{z/4}~ h ~ \textrm{Mpc}^{-1},
 \eeq
while the free-streaming scale is
\beq
k_{\rm fs} \equiv \frac{a^2 H}{u_0} \approx 0.2 ~a^ {-1/2}\frac{300~ \textrm{km/s}}{u_0} ~ h ~ \textrm{Mpc}^{-1},
\eeq
where the last approximation holds during matter domination, $z \gtrsim 0.3$. We see that the free-streaming scale is larger than the non-linear scale for $u_0$ greater than a few hundred km/s, at $z = 0$ and for $u_0 \gtrsim 100$ km/s at $z = 1$. Provided $u_0$ is large enough one can therefore approximate $c_M = 1$ to compute Eq.~\eqref{eq:P_u0-av} on all scales.

\begin{figure*}
\includegraphics[width=0.47\textwidth]{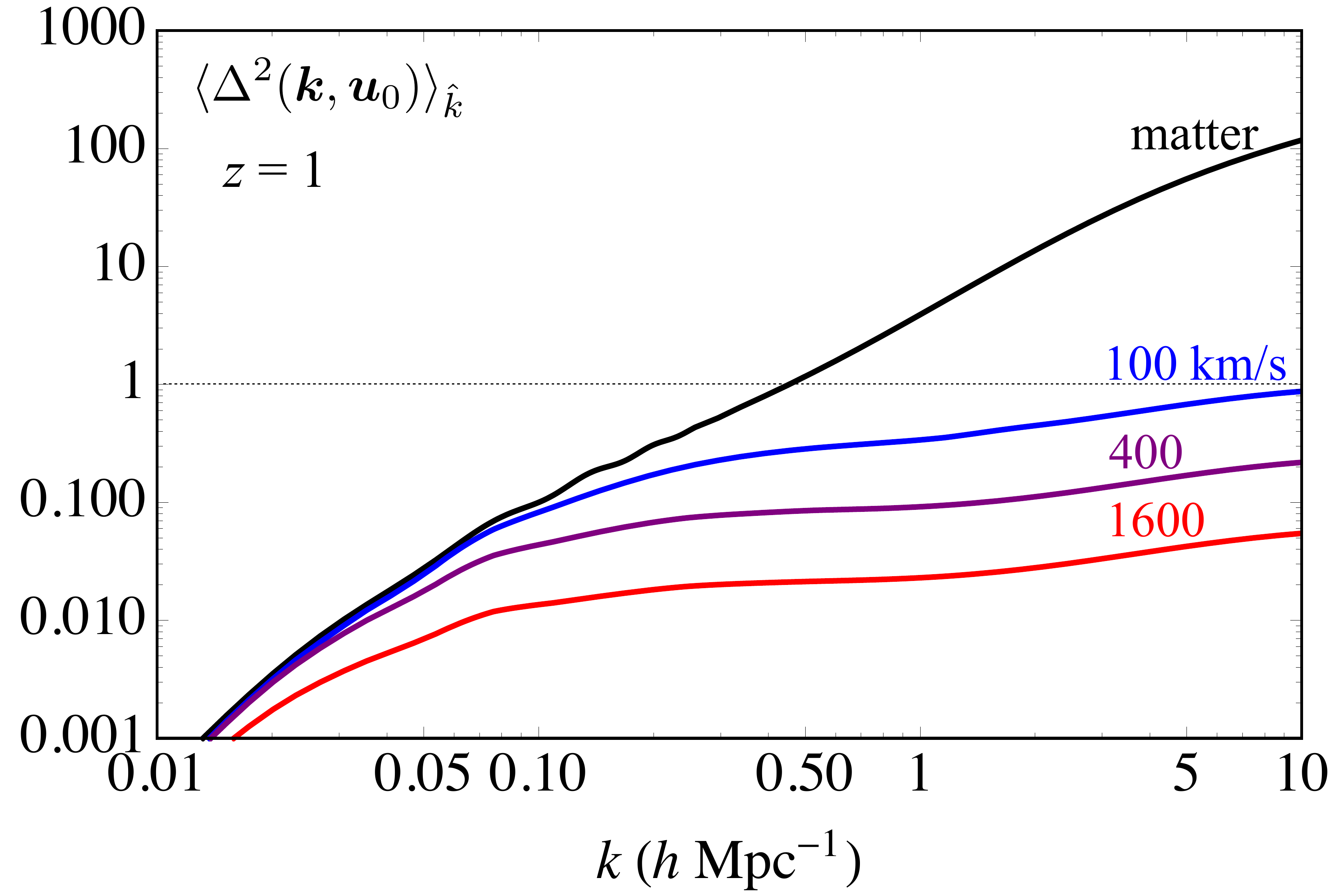}
\includegraphics[width=0.47\textwidth]{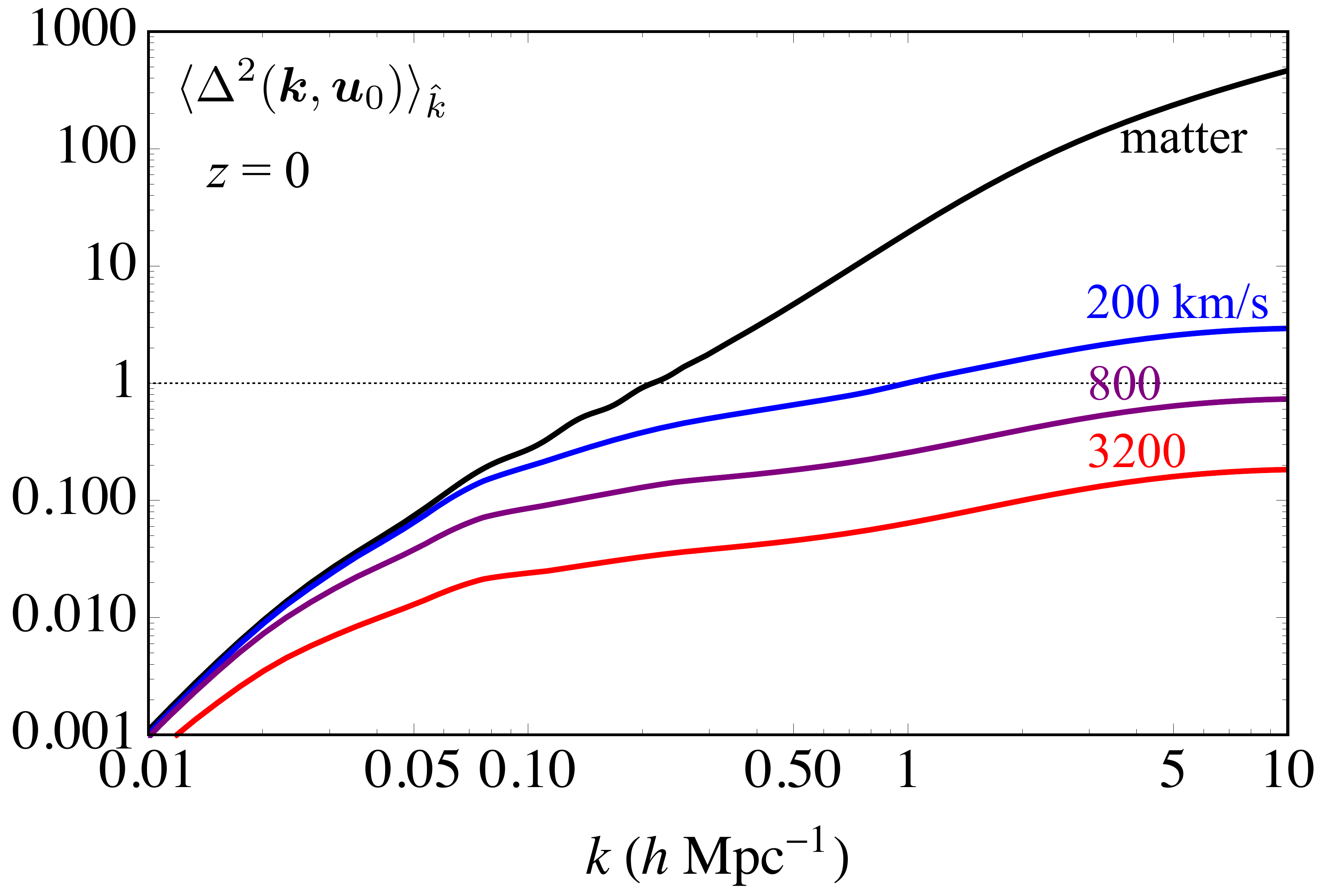}
\caption{Angle-averaged power per logarithmic $k$-interval, in the linear response approximation, for an unperturbed phase-space density $f^0(\bs{u}) = \delta_{\rm D}(\bs{u}- \bs{u}_0)$. The curves are labeled by the unperturbed velocity $u_0$. At redshift 1 (left panel), we see that particles with unperturbed velocities as low as $\sim 100$ km/s do not strongly cluster on all scales. At redshift 0 (right panel), particles with unperturbed velocities $u_0 \gtrsim 800$ km/s do not strongly cluster on all scales.}
\label{fig:halofitvshell}
\end{figure*}

We show $\langle \Delta^2_{\bs{u}_0}(s, \bs{k})\rangle_{\hat{k}}$ in Fig.~\ref{fig:halofitvshell}, at $z = 0$ and 1. We see that at $z = 1$, it is less than unity on all scales for $u_0 \gtrsim 100$ km/s. On the other hand, at $z= 0$, only neutrinos faster than $\sim 800$ km/s are statistically unclustered. These unperturbed velocities therefore provide thresholds above which we expect the LRA to be accurate.

\subsubsection{Single velocity shell}

To confirm our estimates from the single velocity bin case, we now consider a single velocity \emph{shell}, so that we can compare our results to existing numerical results. We therefore consider the following isotropic unperturbed distribution, which is the average of the single-bin distribution over the direction of $\bs{u}_0$:
\beq
f^0_{u_0}(u) \equiv \langle f_{\bs{u}_0}^0 (\bs{u}) \rangle_{\hat{u}_0} = \frac1{4 \pi u_0^2} \delta_{\rm D}(u- u_0).
\eeq
where $\delta_{\rm D}$ is the Dirac delta. In the LRA, the resulting overdensity is (neglecting the initial condition piece)
\barr
&&\delta_{u_0}(\bs{k}) = \langle \delta_{\bs{u}_0}(\bs{k}) \rangle_{\hat{u}_0} \nonumber\\
&&= \frac32 H_0^2 \Omega_M \int_{s_i}^s ds' (s - s') a' j_0[u_0 k (s - s')] \delta_{\rm M}(s', \bs{k}).~~~~~ \label{eq:delta-shell}
\earr
We denote the corresponding dimensionless power spectrum by $\Delta^2_{u_0}(k)$. Let us first point out that the inequality $|\delta_{\bs{u}_0} - \langle \delta_{\bs{u}_0}(\bs{k}) \rangle_{\hat{u}_0}|^2 \geq 0$ implies $\Delta^2_{u_0} \leq \langle \Delta^2_{\bs{u}_0} \rangle_{\hat{u}_0}$. Imposing $\langle \Delta^2_{\bs{u}_0} \rangle_{\hat{u}_0} \leq 1$ is more stringent than imposing $\Delta^2_{u_0} \leq 1$. We will see that the latter inequality can be amply satisfied even in regimes where the LRA fails.

Before computing the power spectrum, it is useful to simplify Eq.~\eqref{eq:delta-shell} further. In the linear regime, the matter overdensity is proportional to the direction-independent (but $k$-dependent) growth rate, so that
\beq
\delta_M(s', \bs{k}) \approx \left(\frac{P_M(s', k)}{P_M(s, k)}\right)^{1/2} \delta_M(s, \bs{k}). \label{eq:phases}
\eeq
On scales smaller than the free-streaming scale, the rapidly oscillating spherical Bessel function selects mostly times $s' \approx s$ in the integral of Eq.~\eqref{eq:delta-shell}. Hence, once again, provided $k_{\rm fs} \lesssim k_{\rm nl}$, the approximation Eq.~\eqref{eq:phases} leads to an accurate $\delta_{u_0}$ on all scales. While we made a similar argument to compute $\langle \Delta^2_{\bs{u}_0} \rangle_{\hat{u}_0}$ earlier, we stress that here, the argument applies to the overdensity field itself, rather than its power spectrum.

Inserting the approximation \eqref{eq:phases} into Eq.~\eqref{eq:delta-shell}, we find the power spectrum
\barr
\Delta^2_{u_0}(s, k)&\approx& \left(\frac32 \frac{H_0^2 \Omega_M}{k u_0}\right)^2\nonumber\\
&\times& \left[\int_{s_i}^s ds' a' \sin[k u_0 (s - s')] \Delta_M(s', k) \right]^2. ~~~~~
\earr
We compare this approximation with the shot noise reduced $N$-body simulations of \cite{Banerjee_2018}, hereafter B18. In this work, massive neutrinos were simulated with the $N$-body method, but with optimized initial conditions in order to reduce shot noise: instead of assigning neutrinos with a random velocity direction at each grid point, $12 ~N_{\rm side}^2$ particles are assigned to each grid point, with an isotropic velocity distribution using the \textsc{healpix} algorithm \citep{Healpix}. Specifically, we compare to the simulation \texttt{LU\_SH10\_NS2} of B18, for which $48 \times 128^3$ neutrino particles were used per velocity shell, for 10 shells ranging from 495 km/s to 5773 km/s, corresponding to a total of approximately $1000^3$ neutrino particles.

\begin{figure}
  \includegraphics[width=0.45\textwidth]{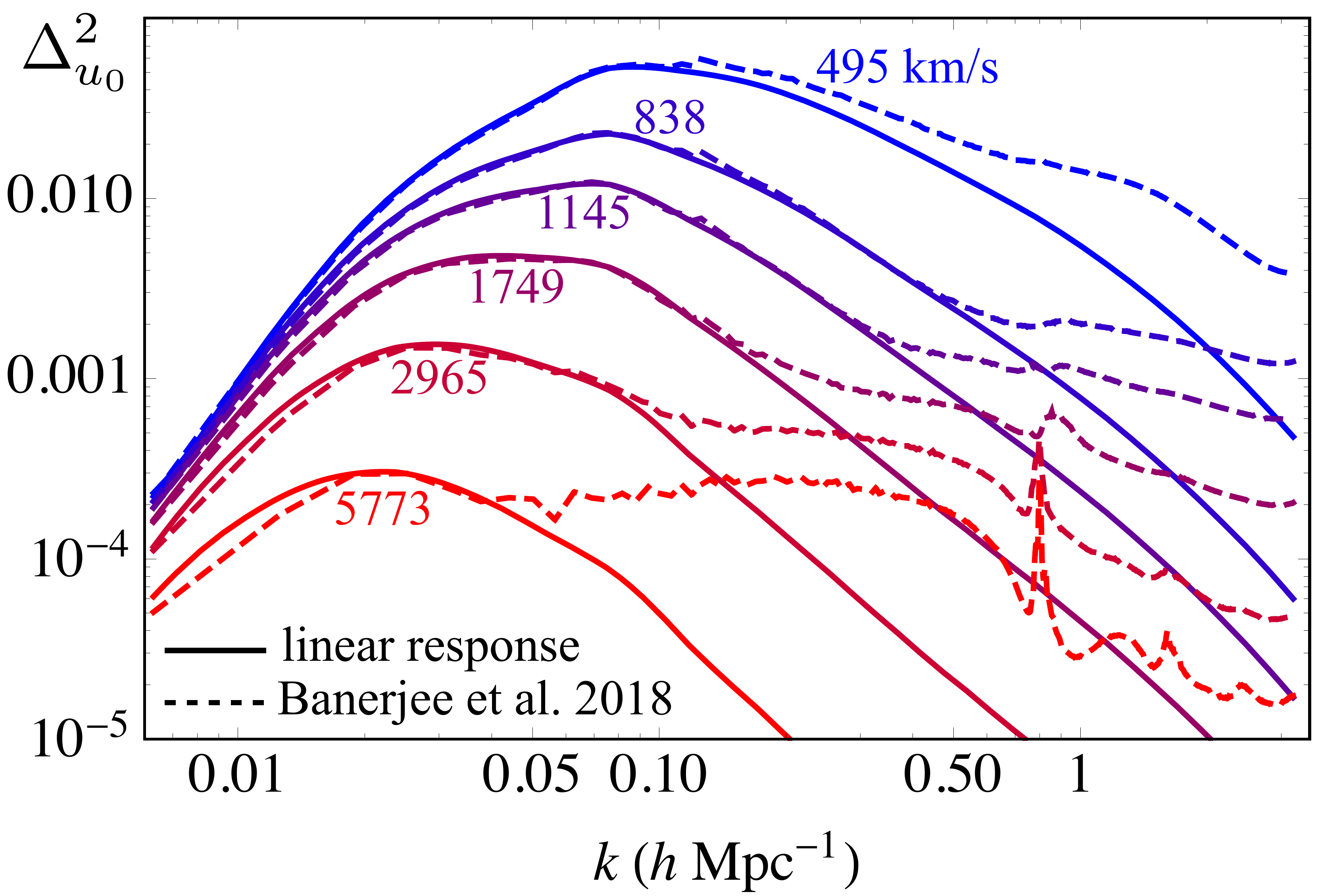}
  \caption{Power per logarithmic $k$-interval in different velocity shells at $z = 0$, in the linear response approximation (LRA, solid), and from the simulations of \protect \cite{Banerjee_2018} (dashed). The agreement is good for $u_0 \gtrsim 800$ km/s, except for departures at small scales due to residual shot noise in the particle simulations. Only for the lowest velocity shell shown does the LRA noticeably underestimate clustering.
  }
  \label{fig:simvshell}
\end{figure}

The comparison with B18 is shown in Fig.~\ref{fig:simvshell}. On large enough scales, we find that the power spectra computed with both methods agree very well (up to sample-variance deviations on scales comparable to the box size of B18). For unperturbed velocities $u_0 \gtrsim 800$ km/s, the results of B18 show clear evidence of residual shot noise, and start deviating from ours at increasingly large scales as the velocity is increased. However, they agree very well with our results on all scales where shot noise is subdominant. Only for the smallest velocity shell considered in B18, $u_0 = 495$ km/s, do we find that our results under-estimate the non-linear clustering of neutrinos, by a factor no larger than 2, for scales $k \gtrsim 0.1~h$ Mpc$^{-1}$. This departure occurs despite the fact that $\Delta^2_{u_0}$ is at most $\sim 0.05$, which again emphasizes that the perturbation of a velocity shell is not a good indicator of the validity of the LRA.
The agreement of our results with their numerical counterparts corroborates our previous finding that the LRA is accurate for $u_0 \gtrsim 800$ km/s at $z = 0$. Since the free-streaming scale corresponding to $u_0 \gtrsim 800$ km/s is well in the linear regime, we are justified a posteriori in using Eq.~\eqref{eq:phases}. We will check explicitly later on that neutrinos with $u_0 \gtrsim 800$ km/s are very well correlated with the matter field, as is implied by this approximation.

\subsection{Application to fast massive neutrinos}

Having established the regime of validity of the LRA, we may now apply it to arbitrary phase-space densities, provided they are restricted to velocities $u \geq v_{\rm crit}$. In addition, since in the regime of validity of the LRA, the free-streaming scale is larger than the non-linear scale, we may also safely make the approximation \eqref{eq:phases} in the integral \eqref{eq:delta-phi}.

\begin{figure}
\includegraphics[width=0.43\textwidth]{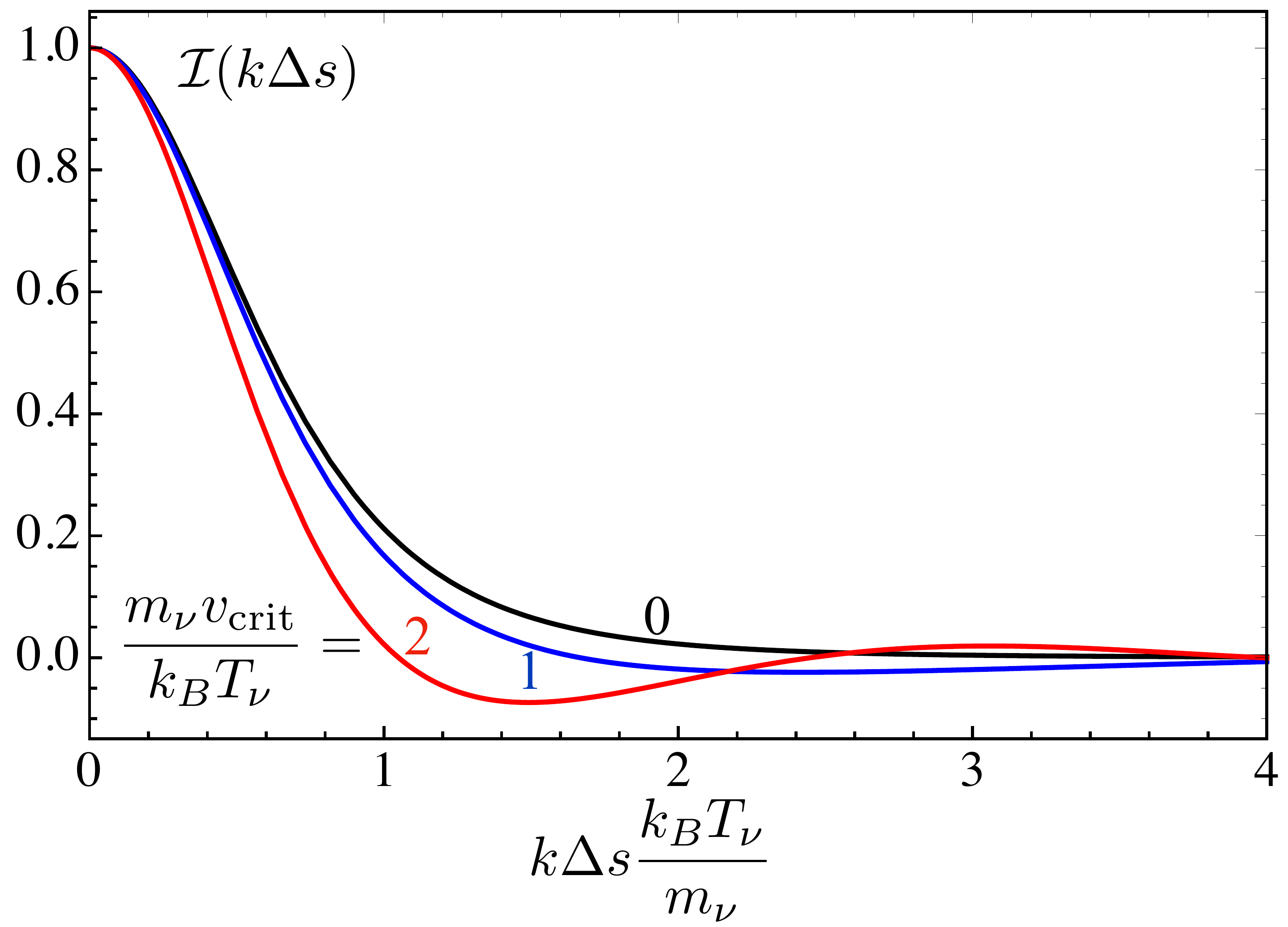}
\caption{Kernel $\mathcal{I}(k \Delta s)$ defined in Eqs.~\eqref{eq:delta-phi}, \eqref{eq:I(k)} for a truncated Fermi-Dirac distribution restricted to $u \geq v_{\rm crit}$, for several values of $m_\nu v_{\rm crit}/k_B T_\nu$. }
\label{fig:kernel}
\end{figure}

Let us apply the LRA to a truncated Fermi-Dirac distribution, which describes the unperturbed velocity distribution of neutrinos faster than $v_{\rm crit}$:
\begin{equation}
f^0(u) \propto \Theta(u - v_{\rm crit}) \left(\rme^{m_\nu u/ \left(k_\mathrm{B} T_\nu\right)} + 1 \right)^{-1},
\end{equation}
where $\Theta$ is the Heaviside function (or an infinitesimally smoothed version of it, so that $f^0$ is differentiable), $m_\nu$ is the mass of a single neutrino or antineutrino, $k_\mathrm{B}$ is the Boltzmann constant in eV/K and $T_\nu \approx 1.97$ K is the temperature of the cosmic neutrino background. The normalization is again such that $\int d^3 u f^0(u) = 1$. The kernel in Eq.~\eqref{eq:delta-phi} is therefore given by
\barr
\mathcal{I}(\kappa) = \frac{\int_{q_c}^{\infty} dq~ j_0\left(\kappa \frac{k_\mathrm{B} T_\nu}{m_\nu}q\right)~ q^2 /(\rme^q + 1) }{\int_{q_c}^{\infty} dq ~q^2/(\rme^q + 1)}, \ \ \ q_c \equiv \frac{m_\nu v_{\rm crit}}{k_\mathrm{B} T_\nu}.
\earr
We use the following asymptotic expansion for the integral:
\begin{align}
 \int^\infty_{q_\mathrm{c}} \frac{j_0(qX)}{e^q + 1} q^2 dq &= - \sum^{\infty}_{n=1} (-1)^n \frac{\rme^{-n q_\mathrm{c}}}{(n^2+X^2)^2} I_n(q_\mathrm{c},X),\\
 I_n(q_\mathrm{c},X) &= \left(n^2 + n^3 q_\mathrm{c} + n q_\mathrm{c} X^2 - X^2\right) \frac{\sin(q_\mathrm{c} X)}{X} \nonumber \\
 &+ \left(2n + n^2 q_\mathrm{c} + q_\mathrm{c} X^2\right) \cos(q_\mathrm{c} X),
\end{align}
We have checked that this expansion is converged to within an absolute error of $< 10^{-4}$ for $n_\mathrm{max} = 20$. We show the kernel $\mathcal{I}$ for different values of $v_{\rm crit}$ in Fig.~\ref{fig:kernel}.

\section{Hybrid method} \label{sec:hybrid}

\subsection{Basic idea and default parameters}

\begin{figure}
\includegraphics[width=0.45\textwidth]{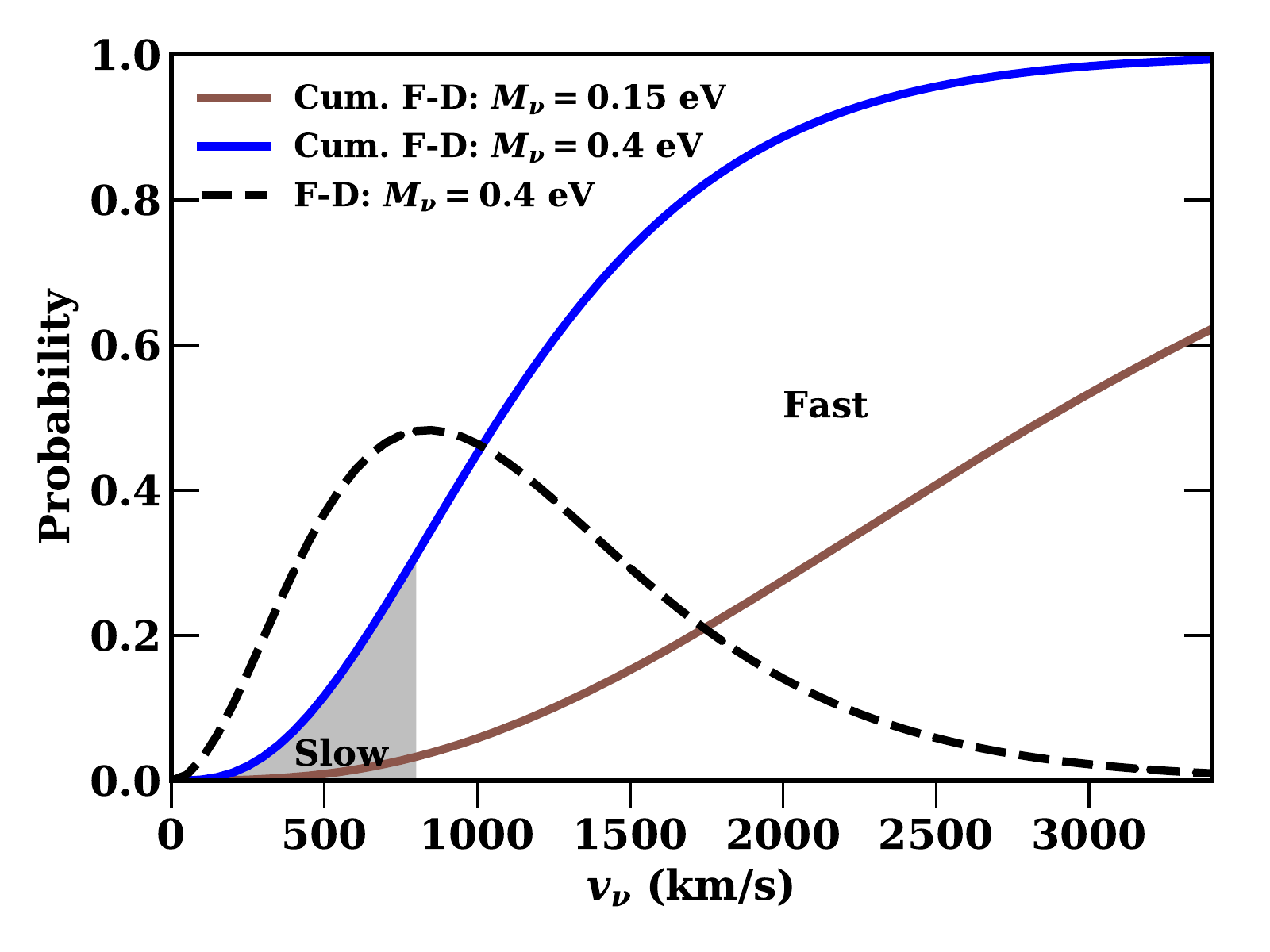}
  \caption{The integrated Fermi-Dirac distribution, showing the cumulative probability for neutrinos to have an unperturbed velocity less than $v_\nu$ at $z=0$ for total neutrino mass $M_\nu = 0.4$ and $M_\nu = 0.15$ eV.
  The grey shaded region shows the neutrino density followed by particles in our hybrid method, which is independent of neutrino mass.
  }
  \label{fig:fddistribution}
\end{figure}

We split neutrinos into a ``slow" and a ``fast" component, depending on their \emph{unperturbed} velocity relative to some critical velocity $v_{\rm crit}$, as illustrated in Fig.~\ref{fig:fddistribution}. We emphasize that this split is performed in the unperturbed velocity; acceleration due to passing in and out of potential wells formed by structure may cause ``slow'' particles to acquire a large perturbed velocity, and ``fast'' particles to slow down. The ``slow" and ``fast" labels can be thought of ``Lagrangian'' labels, which allow both the slow and fast components to independently satisfy the collisionless Boltzmann equation\footnote{If the split between ``slow" and ``fast" had been made instead in terms of the perturbed or ``Eulerian" velocity, one would have to solve for two coupled collisional Boltzmann equations to account for the flux of particles across the boundary.}. We are free to solve these two equations with different techniques: the LRA for fast particles and the $N$-body method for slow particles. This allows us to focus the computationally expensive $N$-body technique on the fraction of phase-space where it is truly needed. Note that at fixed critical velocity, the lighter the neutrino, the smaller the fraction of ``slow" particles.

We could in principle set a time-dependent velocity boundary $v_{\rm crit}(z)$. However, this would require constantly converting fast particles into slow particles, since the range of validity of the LRA decreases with time, as neutrinos redshift and gravitational potentials deepen. Instead, we use the LRA for \emph{all} neutrinos down to a redshift $z_\nu$, after which we use a fixed $v_{\rm crit}$ down to $z = 0$. Our hybrid method thus has two adjustable parameters, $z_\nu$ and $v_{\rm crit}$. The choice of $z_\nu$ must be such that the fraction of neutrinos that are not accurately described by the LRA at $z \geq z_\nu$ is sufficiently small that it has no significant impact on any observables. The choice of $v_{\rm crit}$ must be such that neutrinos with $u_0 \geq v_{\rm crit}$ are accurately described by the LRA until the present.

We saw in Section \ref{sec:single-bin} that we expect the LRA to be accurate for $u_0 \gtrsim 100$ km/s at $z \geq 1$. For a relativistic Fermi-Dirac distribution with temperature $T_\nu = 1.95$ K, the fraction of neutrinos with unperturbed velocity less than $100$ km/s is
\beq
P(u < 100 ~\textrm{km/s}) \approx 6.7 \times 10^{-4} \left(\frac{m_{\nu}}{0.1~\textrm{eV}}\right)^3.
\eeq
Therefore, by treating \emph{all} neutrinos with the linear response approximation at $z \geq 1$, we underestimate the clustering of at most $\sim 0.1\%$ of the neutrinos, given current upper bounds to their masses.
Even if these slow neutrinos clustered as strongly as the CDM (a substantial over-estimate), mis-estimating their clustering would still have a negligible effect on the total matter power on any scale. Indeed, the fractional error on $P_\mathrm{m}(k)$ is bounded by $\sim 10^{-3} \times 10 f_{\nu}$, where $f_\nu \approx 0.02 (M_\nu/0.3 \textrm{eV})$ is the fraction of matter in neutrinos, and the factor of $10 f_\nu$ accounts for the cumulative effect of neutrinos on non-linear scales (slightly stronger than the well-known $8 f_\nu$ suppression on linear scales). We see that this error should be very small, so we deem it safe to use the method of AHB13 (the LRA applied to \emph{all} neutrinos) for $z \geq 1$, i.e.~to set $z_\nu = 1$.\footnote{Note that \cite{Banerjee_2018} do not resolve neutrino velocities below several hundred km/s at all in their default setup, even for neutrino masses as low as $M_\nu = 0.15$ eV.}

We also saw in Section \ref{sec:validity} that the LRA is accurate at $z = 0$ for $u_0 \gtrsim 800$ km/s. To be conservative, we use a default neutrino critical velocity of $850$ km/s for $0 \leq z \leq z_\nu$. For reference, the fraction of slow neutrinos is $P(u < 850 ~\textrm{km/s}) \approx [0.04, 0.20, 0.42]$, respectively, for  $m_{\nu} = [0.05, 0.1, 0.15]$ eV.

In addition to the LRA, we use the approximation \eqref{eq:phases} in Eq.~\eqref{eq:delta-phi}, whose regime of validity we have shown to be broader than that of the LRA. This means that to compute the density field of fast neutrinos at some redshift $z$, we only require the 3-dimensional matter density field at $z$, as well as the one-dimensional matter power spectrum as a function of time. In practice, we store the latter at even intervals $\Delta a = 0.01$, and then interpolate it when required\footnote{The code actually provides the CDM power spectrum. We estimate the matter power spectrum by adding the neutrino component from the previous timestep assuming it is completely correlated, i.e. $P_m^{1/2} = (1 - f_\nu)P_{\rm dm}^{1/2} + f_\nu P_\nu^{1/2}$.
It would be possible to iterate this procedure, but in practice it is always immediately converged to a high degree of accuracy, see Appendix B of AHB13.}.

To summarize, our default parameters are $z_\nu = 1$ and $v_{\rm crit} = 850$ km/s. These parameters can be adjusted and optimized depending on the desired observable. The standard particle technique for all neutrinos corresponds to $z_\nu \rightarrow \infty$ and $v_{\rm crit} \rightarrow \infty$. The method of AHB13 corresponds to $z_\nu \rightarrow 0$ or $v_{\rm crit} \rightarrow 0$.

Finally, just like in AHB13, we solve for the evolution of neutrinos and CDM simultaneously, and accounting for their mutual gravitational interaction: at each time step, we update the velocities and positions, hence overdensity of CDM and slow-neutrino particles with the $N$-body algorithm, and we use the LRA to compute the overdensity of fast-moving neutrinos.

\subsection{Initialization of the ``slow'' neutrinos}

We generate slow neutrino particles at the initial simulation redshift, $z_i = 99$, and follow them as tracer particles until $z_\nu$. Their trajectories are thus computed using the $N$-body technique in the total matter potential, comprising CDM and linear response neutrinos. However, they are not used to compute the potential until after $z_\nu$\footnote{Note that the hybrid method of \cite{Brandbyge_2010} creates neutrino particles dynamically during the simulation, rather than initially treating them as tracers.}.

We could in principle generate slow neutrino particles at the cutoff redshift $z_\nu$, using the LRA to compute the initial neutrino overdensities and bulk velocities, as a function of their thermal velocity $\bs{u}_0$. In practice however, there are several advantages to adding neutrino particles at the initial time of the simulation. The first is implementation simplicity; computing the perturbed velocities to sufficient accuracy is not trivial, whereas the unperturbed velocities are simply a truncated Fermi-Dirac distribution. The second is that by introducing the neutrino particles as tracers we naturally include any higher order (non-Gaussian) neutrino correlations arising from growth in the CDM at $z > z_\nu$.

The increased particle load still imposes some numerical overhead over the pure linear response method at early times, but we found this to be relatively small. In our tests, before $z_\nu$, hybrid particle simulations were about $35\%$ slower than pure LRA simulations, while a pure particle simulation was about $150\%$ slower with the same particle load.
The numerical overhead is thus substantially reduced for the hybrid method over pure particle neutrino simulations. This is mostly because the hybrid method uses neutrino particles with a lower average velocity and thus longer timesteps.

Simulating slow neutrino trajectories from $z_i = 99$ also allows us to start from completely homogeneous initial conditions, i.e. neglect any initial overdensities and bulk velocities. We verified explicitly that our results at all redshifts were unchanged for a simulation where our slow-moving neutrinos had an initial clustering matching the transfer function of the CDM, a conservative over-estimate of their true initial clustering.

For ease of comparison with particle simulations, the simulations presented in this paper assume that all three neutrino species are degenerate. Future hybrid simulations including a neutrino hierarchy would generate particles only for the most massive neutrino species, as for particle simulations. Note that neutrino masses small enough that the neutrino hierarchy is important, $M_\nu < 0.15$~eV, have $\lesssim 4$\% of neutrino initial velocities $< 850$ km/s. AHB13 showed that the linear response method is able to accurately model the neutrino power spectrum at these masses.

\section{Simulations} \label{sec:simulations}

\subsection{Improvements to the particle neutrino implementation}
\label{sec:partnuimprovements}

Our simulations were run with MP-\gadget\footnote{\url{https://github.com/rainwoodman/MP-Gadget/}, rev. fcb861eb}. Our implementation of particle neutrinos includes several modifications to improve accuracy.
In \cite{Bird_2012} we disabled the short-range tree force for the neutrino particles, leaving them affected only by the long-range particle-mesh force. Since the purpose of this work is to follow the non-linear small-scale evolution of the neutrinos accurately, we enable the short-range tree force in this work. The timestep for the short-range force of the neutrino particles is set based on their acceleration, with the same criterion that is used for CDM \citep{Springel_2005}. However, the time between computation of long-range particle-mesh forces is computed without reference to the neutrino particles, as in \cite{Viel_2010} and \cite{Bird_2012}. In \gadget this timestep is set by particle displacements, so that no particle moves more than a fraction of a grid cell in a single timestep. The inclusion of neutrinos would make it extremely small.

We found that the large velocity of neutrino particles caused severe numerical issues in \gadget, manifesting as frequent hangs when walking the short-range gravitational force tree. We traced these problems to an optimisation introduced in \gadget-3. In order to avoid the overhead of regenerating the force tree every timestep, the force tree persists over short timesteps. Particle motion is accounted for by moving tree nodes according to the average velocity of particles within each node. Neutrino particles have large random thermal velocities, which cause them to frequently move from one force node to another. It is thus not sufficient to account for particle motion by moving the node of the force tree, and one must instead rebuild the force tree from scratch every timestep. We found that a more frequent rebuilding of the force tree improves the accuracy of the simulation even without neutrino particles, and thus we have made rebuilding the force tree each timestep the default behaviour in MP-\gadget.

We implemented a number of optimizations in the tree build code to avoid it dominating the cost of the simulation.
In particular, we implemented a cache of the last location a particle was added to the tree. Our tree-build code first attempts to add a new particle to the cached location, and only walks the partially-built tree from the top if this fails. This doubled tree building speed. Gadget-3 pre-computes Morton keys for each particle, using them as a fast hash to know where in the tree each particle should be placed. However, modern processors can use vector instructions to place a particle quickly without the need for this hash, which merely pollutes the processor cache. Removing this mis-optimisation again doubled tree building speed. A few tens of percent improvement was achieved using micro-optimisations. Finally, scalability was improved by threading the treebuild with OpenMP. With these optimisations, in our hybrid simulation the tree build stage was $11\%$ of the total walltime.

\subsection{Suite of simulations}

\begin{table}
\begin{center}
\begin{tabular}{|l|c|c|c|c|l|}
\hline
Name & $M_\nu$ (eV) &  $N_\nu^{1/3}$ & $v_\mathrm{crit}$ & $z_\nu$ \\
\hline
\texttt{CDM}    &       0             &             0         & - & -    \\
\texttt{LINRESP-MINNU}    &     0.06 (NH)      & 0         & - &  -  \\
\texttt{PARTICLE-MINNU}    &     0.06 (NH)      & 512      & - &  -  \\
\texttt{LINRESP}   &     0.4            & 0         & - &   - \\
\texttt{PARTICLE}    &     0.4             &   512       & - &   - \\
\texttt{PARTICLE-1024}    &     0.4             &    1024 & - & -    \\
\texttt{HYBRID}    &     0.4             &    512      & 850 & 1\\
\texttt{HYBRID-256}    &     0.4             &    256       & 850 &1 \\
\texttt{HYBRID-z4}    &     0.4             &    512       & 850 & 4  \\
\texttt{HYBRID-v750}  &     0.4             &    512       & 750 & 1\\
\texttt{HYBRID-v1000}   &     0.4             &    512       & 1000 & 1\\
\texttt{HYBRID-v5000}   &     0.4             &   512       & 5000 & 1 \\
\hline
\end{tabular}
\end{center}
\caption{Table of simulations performed. All simulations have a box of $300$ Mpc/h
and $512^3$ CDM particles. NH denotes a normal hierarchy for the neutrino masses.
For $M_\nu = 0.4$ eV, the fraction of neutrinos slower than critical velocities of $750$, $850$, $1000$, $5000$ km/s is $0.276$, $0.346$, $0.451$ and $1$ respectively.}
\label{tab:simulations}
\end{table}

We ran a suite of simulations listed in Table \ref{tab:simulations}, all using a box size of $300$ comoving Mpc/$h$ and $512^3$ CDM particles. In all cases, CDM particles are initialised using a linear transfer function generated at $z=99$ with massive neutrinos using CAMB \citep{CAMB_neutrinos} and the Zel'dovich approximation \citep{Zeldovich_1970}. To minimize the effect of sample variance in our comparisons, all simulations use the same unperturbed Gaussian field in their initial conditions.

Our suite includes a simulation with massless neutrinos (\texttt{CDM}) and simulations including massive neutrinos with three different methods: pure linear response method as in AHB13 (prefix \texttt{LINRESP}), pure particle method (prefix \texttt{PARTICLE}) and hybrid method (prefix \texttt{HYBRID}). The simulations \texttt{LINRESP-MINNU} and \texttt{PARTICLE-MINNU} use the minimal neutrino mass allowed by oscillation experiments, $M_\nu = 0.06$ eV, and assume a normal neutrino hierarchy, i.e., specifically, account for two massive neutrinos with masses ($0.009$, $0.05$) eV. Note however that the \texttt{PARTICLE-MINNU} simulation only contains particles corresponding to the largest neutrino mass state, $m_\nu = 0.05$ eV. The $m_\nu  = 0.009$ eV mass state is assumed not to cluster, which is a good approximation as the free-streaming length at $z=0$ is larger than our box size. All other simulations assume three degenerate neutrinos with total mass $M_\nu = 0.4$~eV. This relatively high neutrino mass was chosen to enhance any deviations between different simulation methods.

We found that a neutrino particle simulation with a neutrino particle load of $512^3$ neutrino particles (equal to the CDM particle load) was affected by shot noise and showed a $1\%$ discrepancy with linear theory. Achieving the same accuracy as the LRA for the matter power spectrum with the pure particle simulation requires $1024^3$ neutrino particles for a neutrino mass $M_\nu = 0.4$ eV.

We performed a number of simulations designed to test the sensitivity of our hybrid method to numerical parameters. Our default hybrid neutrino simulation (\texttt{HYBRID}) has $512^3$ neutrino particles (matching the CDM), a neutrino particle switch-on redshift $z_\nu = 1$, and a critical velocity $v_{\rm crit} = 850$ km/s. Our modified simulations considered the effects of lowering the number of neutrino particles to $256^3$ (\texttt{HYBRID-256}), increasing the particle switch-on redshift to $z_\nu = 4$ (\texttt{HYBRID-z4}), and changing the critical velocities to $750$ and $1000$ km/s (\texttt{HYBRID-v750, HYBRID-v1000}). Finally, we ran a hybrid simulation with a critical velocity of $5000$ km/s (\texttt{HYBRID-v5000}), which effectively simulates all neutrinos as particles for $z \leq 1$, as a sanity check of the robustness of our implementation.  The simulation power spectra and scripts for generating most of the figures in this paper can be found at: \url{https://github.com/sbird/hybrid-neutrino}

\section{Results}
\label{sec:results}

\begin{figure*}
\includegraphics[trim={1.5cm 0 3.3cm 0},clip,width=0.29\textwidth]{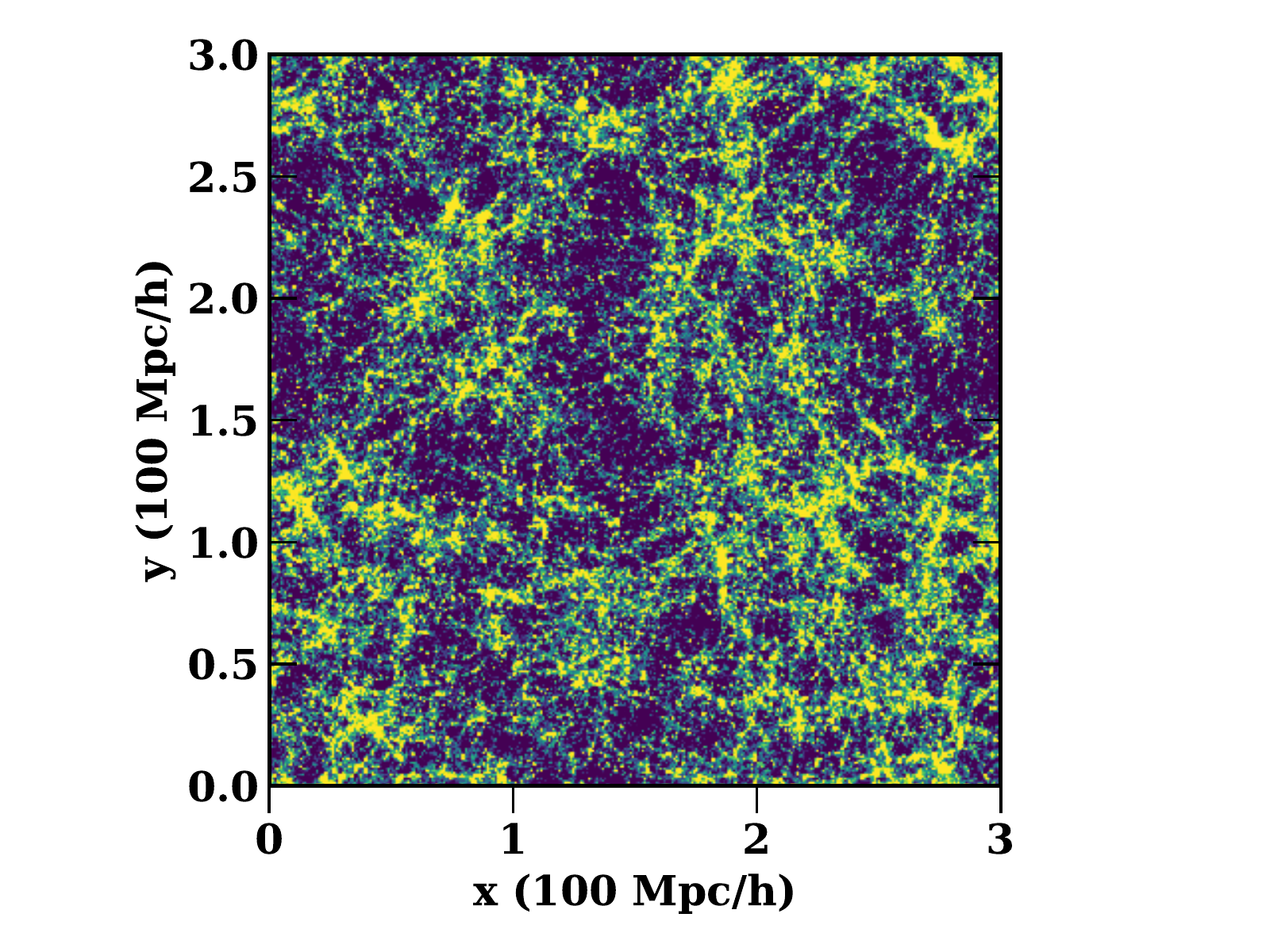}
\includegraphics[trim={1.5cm 0 3.3cm 0},clip, width=0.29\textwidth]{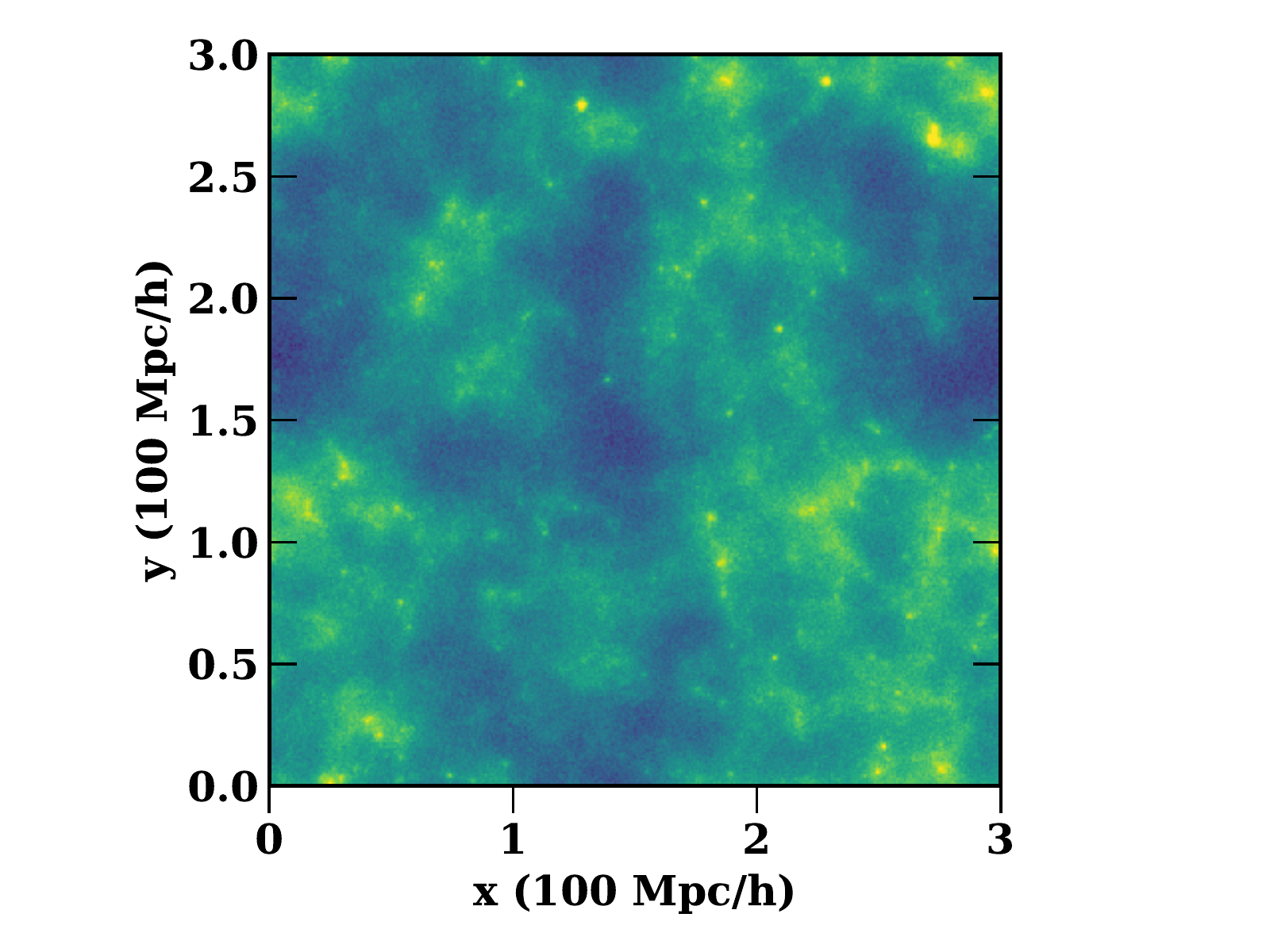}
\includegraphics[trim={1.5cm 0 0.5cm 0},clip, width=0.36\textwidth]{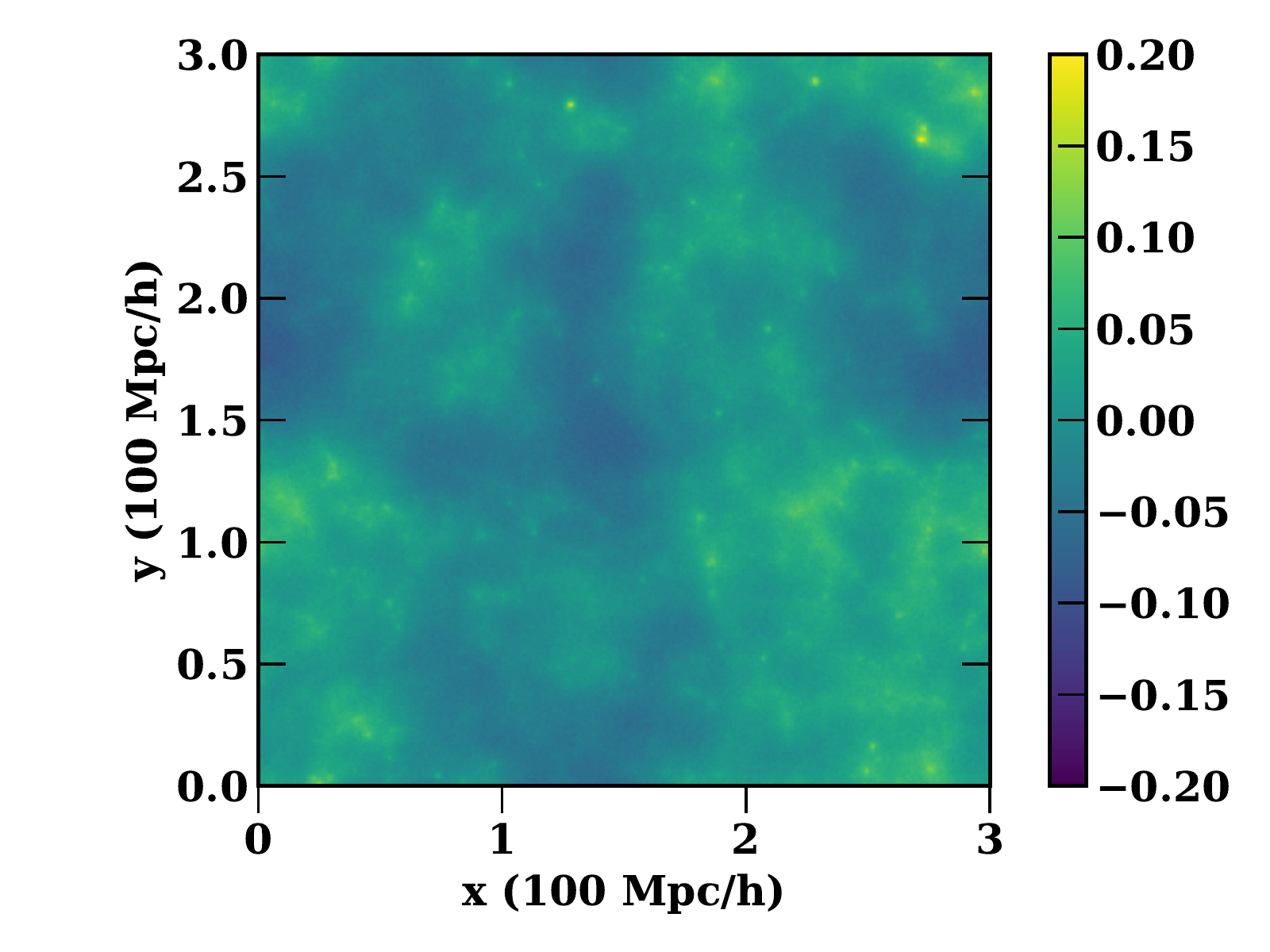}
    \caption{Projected density plots at $z=0$. \emph{Left}: CDM. \emph{Middle}: Neutrino particles from the hybrid simulation, with unperturbed velocity $<850$ (comoving) km/s. \emph{Right}: Neutrino particles from the pure particle simulation, i.e. including neutrinos from all unperturbed velocities. Colours show $\log (1+ \delta)$ in dimensionless units, where $\delta$ is the over-density (computed with respect to the total matter) projected throughout the box.
  The clustering of the slow component of neutrinos followed by the hybrid particles is intermediate between that of the cold dark matter and the clustering of all neutrinos from the particle neutrino simulation. Colour scale has been truncated for the CDM for clarity. Structures have the same positions in all three panels.
  }
  \label{fig:density_plot}
\end{figure*}

\begin{figure*}
\includegraphics[width=0.45\textwidth]{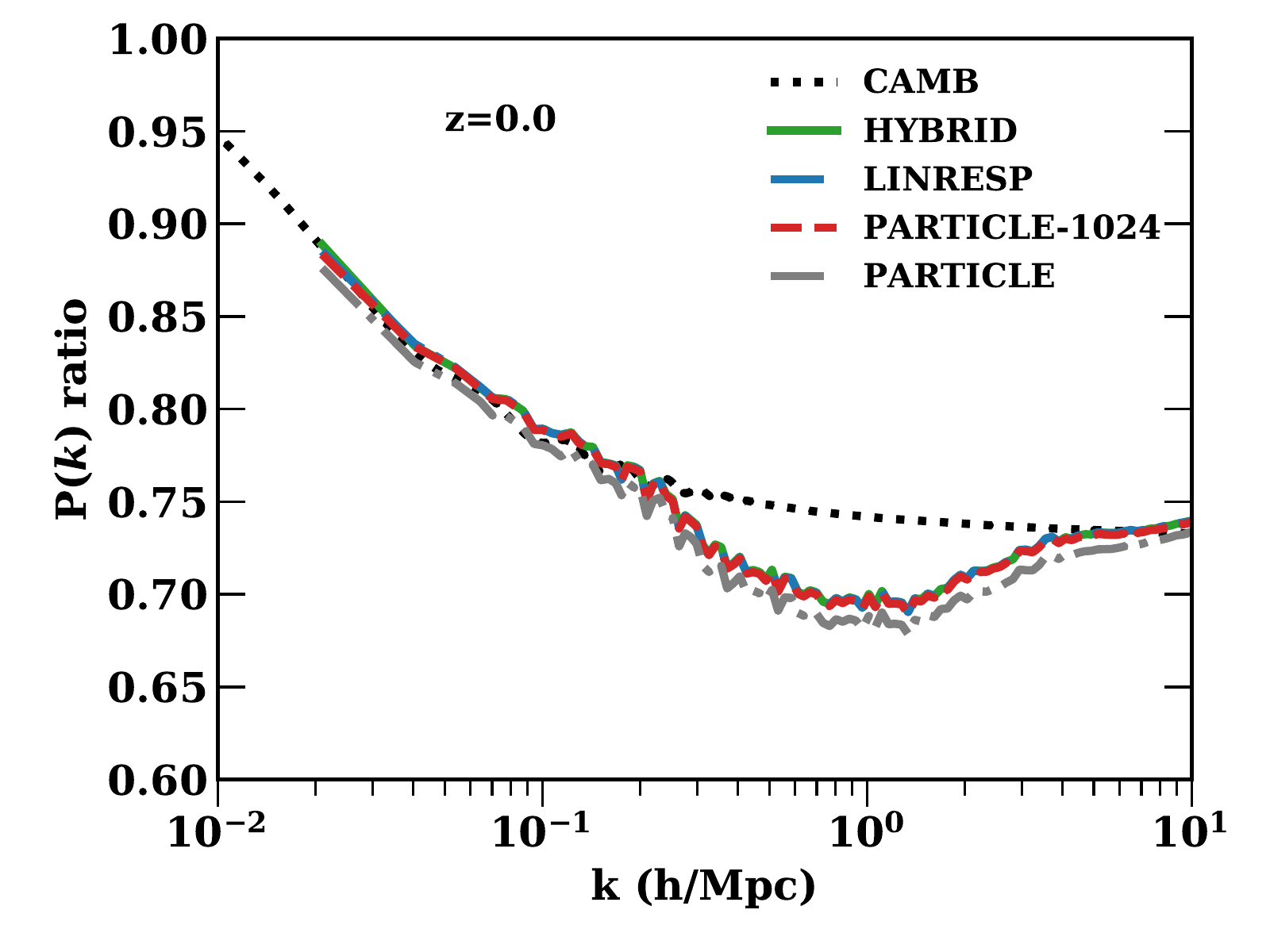}
\includegraphics[width=0.45\textwidth]{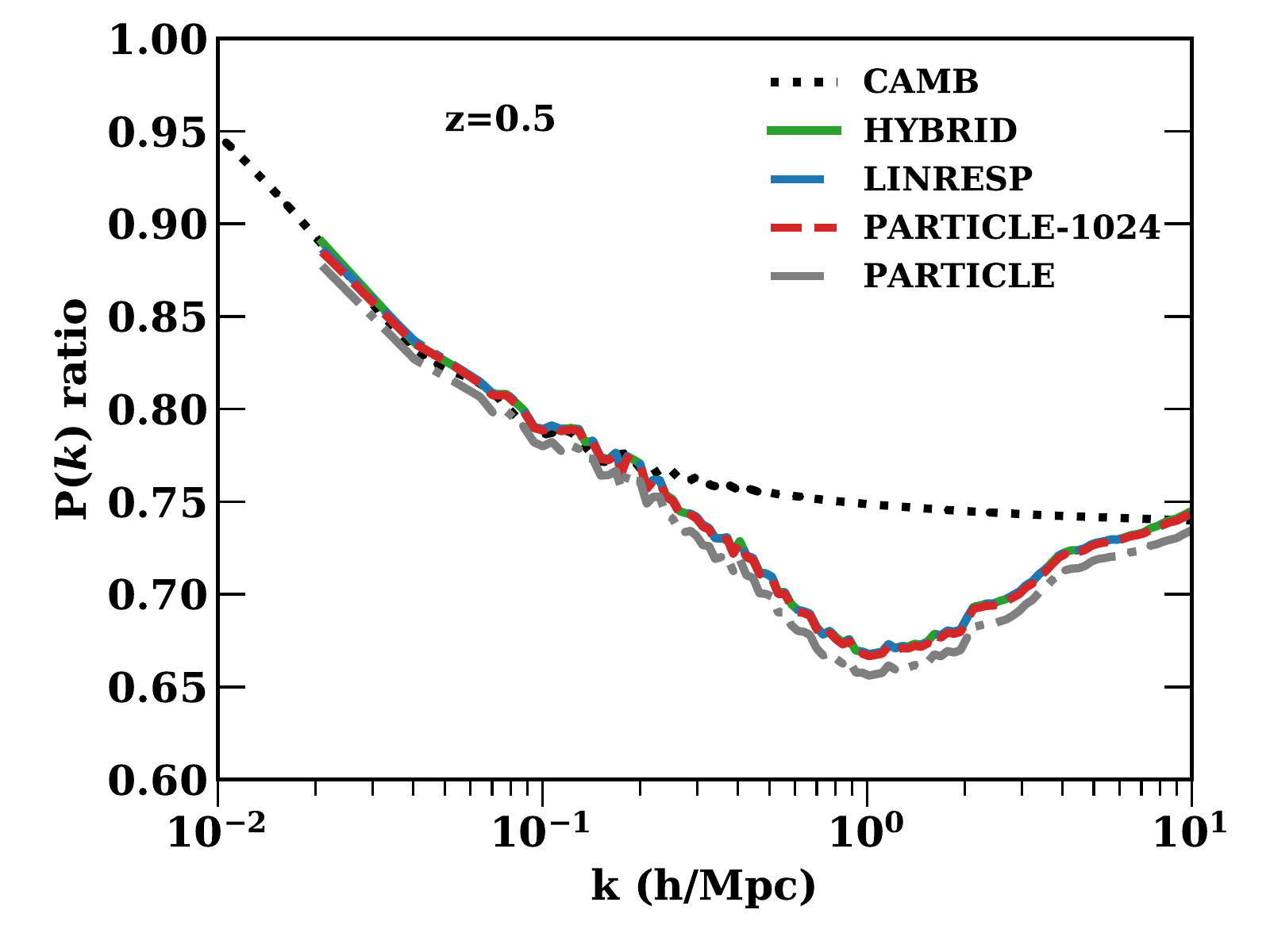}
    \caption{Ratio of matter power spectrum with massive neutrinos ($M_\nu = 0.4$ eV) to matter power spectrum with massless neutrinos. Figure shows hybrid (\texttt{HYBRID}), linear response (\texttt{LINRESP}) and particle methods at $z=0$ (left) and $z=0.5$ (right). Particle simulations with $1024^3$ (\texttt{PARTICLE-1024}) and $512^3$ (\texttt{PARTICLE}) neutrino particles are shown. All simulation methods agree to $\sim 0.1$\%, in agreement with AHB13. The \texttt{PARTICLE} simulation, however, was stronly affected by shot noise. A converged particle neutrino simulation thus required almost an order of magnitude more CPU time than linear response, as discussed in Section~\ref{sec:matterpower}.}
  \label{fig:matter_power}
\end{figure*}

\begin{figure*}
\includegraphics[width=0.45\textwidth]{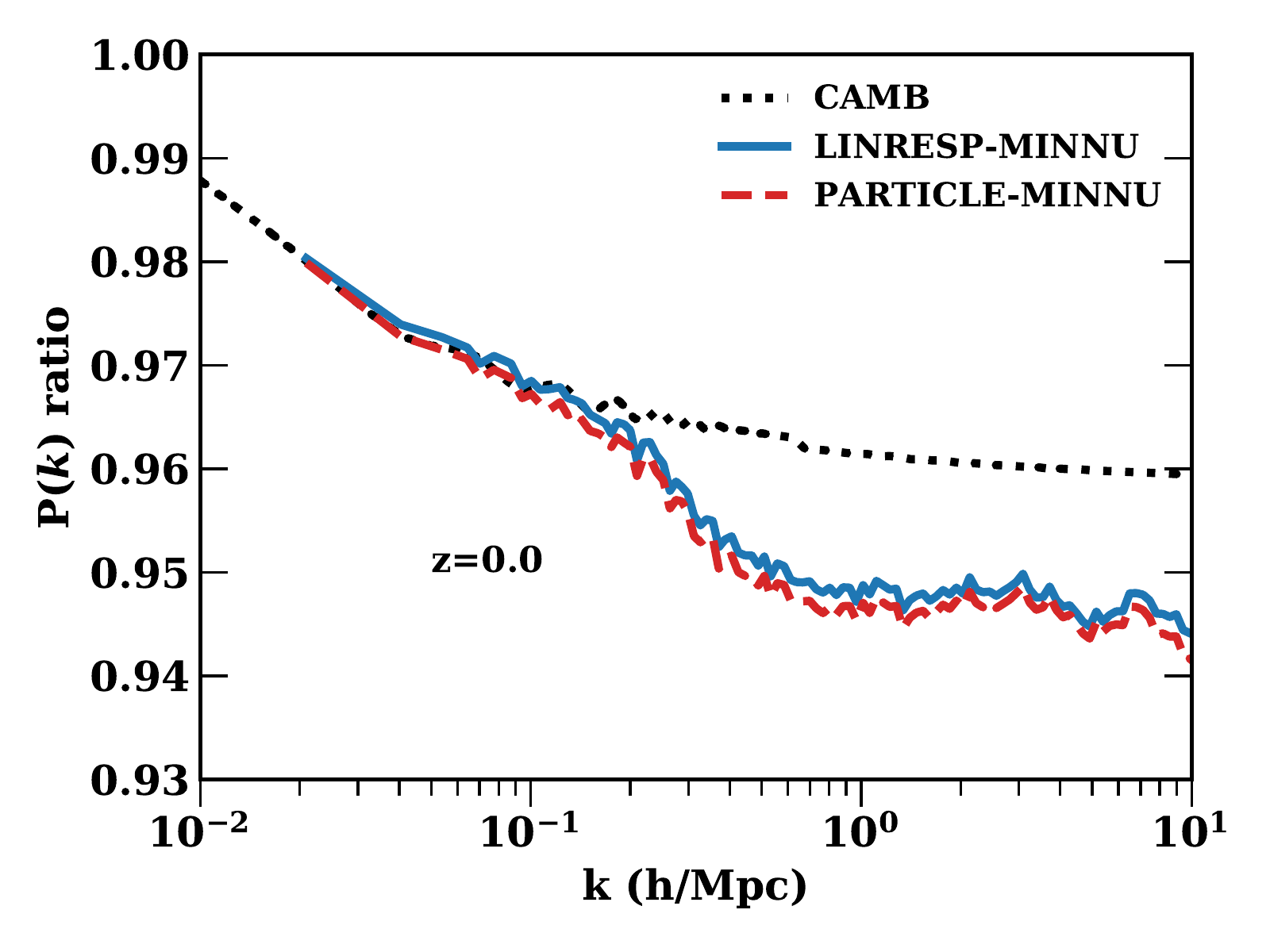}
\includegraphics[width=0.45\textwidth]{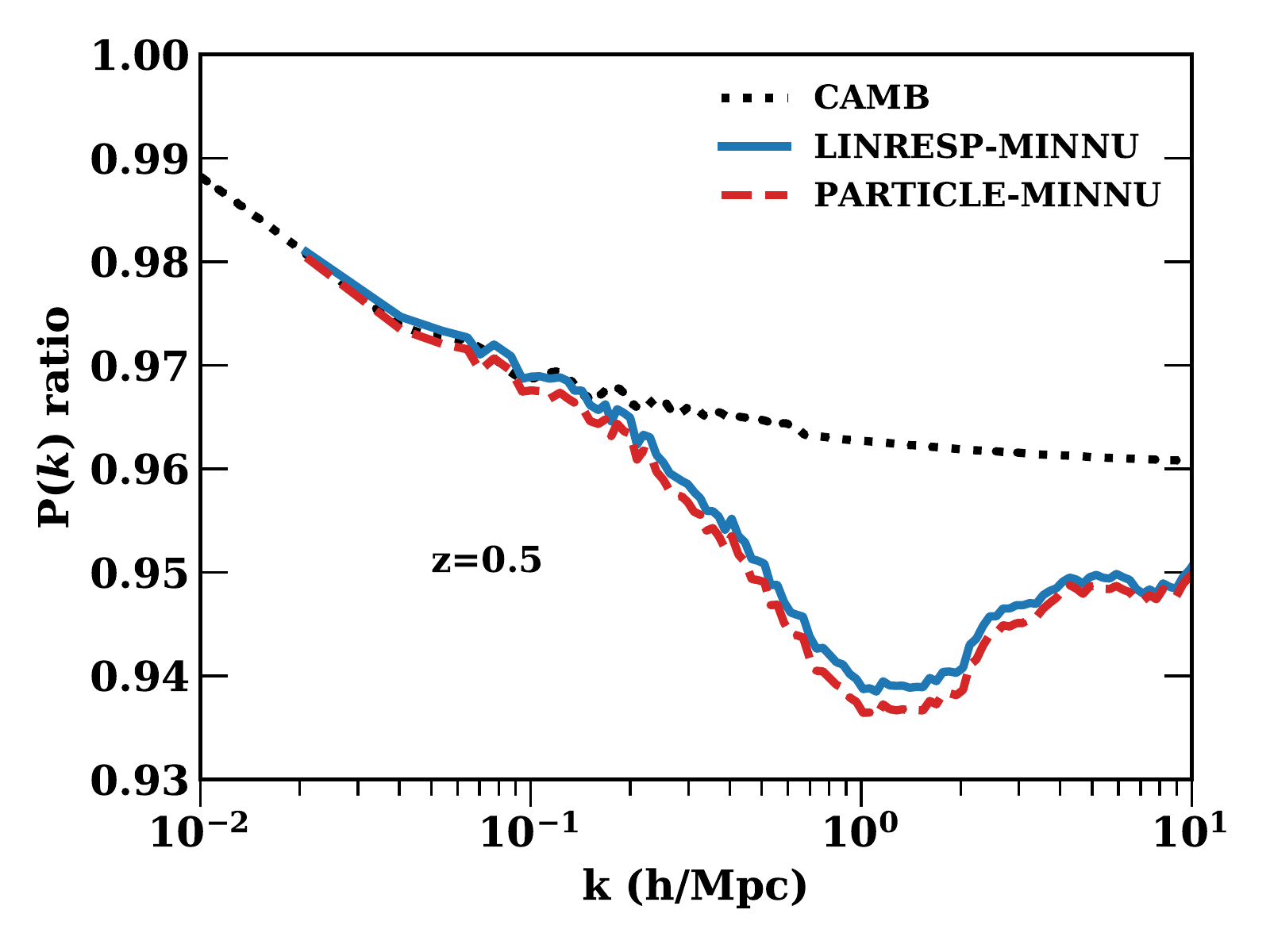}
\caption{Ratio of the matter power spectrum with the minimal neutrino mass ($M_\nu = 0.06$ eV) to the matter power spectrum with massless neutrinos at $z=0$ (left) and $z=0.5$ (right). Shown are simulations using the linear response method (\texttt{LINRESP-MINNU}) and particle method (\texttt{PARTICLE-MINNU}). 
}
\label{fig:minimal_mass}
\end{figure*}

Figure \ref{fig:density_plot} shows projected densities for the CDM and neutrino particles from the HYBRID and PARTICLE simulations, computed using Nbodykit \citep{Hand_2017}, in order to give a visual impression of the clustering of the neutrinos and dark matter. The structure of filaments, halos and voids is identical in all three plots, as the neutrinos and CDM are highly correlated. The neutrino particles in the hybrid simulation, which include only the initially slow tail of the Fermi-Dirac distribution, cluster more than in the purely particle simulation, but in both cases less than the CDM. This matches the expected behaviour given their lower initial thermal velocities.

\subsection{Matter Power}
\label{sec:matterpower}

Figure~\ref{fig:matter_power} shows the $z=0$ and $z=0.5$ matter power spectrum for all three neutrino simulation methods: particle, linear response and hybrid. These three simulations are identical except for the method used to follow neutrino perturbations. In particular, they all use the same cosmological background solution, which includes radiation and massive neutrinos.

Figure~\ref{fig:matter_power} shows the ratio of the matter power spectrum in each simulation with massive neutrinos compared to the matter power spectrum from a pure-CDM massless neutrino simulation with the same total non-relativistic matter density at $z = 0$. We also show the linear theory prediction from CAMB. The hybrid, linear response, and particle methods produce matter power spectra which are indistinguishable by eye, and differ by around $0.1\%$. Given the qualitatively different implementation of massive neutrinos, this suggests that all three methods are well converged. All simulations agree with CAMB at the $1\%$ level on scales which are linear, $k < 0.2$ $h$/Mpc, and small enough to avoid being affected by our finite box size, $k > 0.04$ $h$/Mpc.

The particle neutrino simulation contains $1024^3$ neutrino particles, $8\times$ the number of CDM particles. This high particle load increased the computational cost by a factor of $\sim 10$ (note that our simulation includes the short-range gravitational force for the neutrinos), and memory requirements by a factor of $\sim 8$. We also show in Figure~\ref{fig:matter_power} a particle simulation with $512^3$ neutrino particles, which underestimates the growth of the matter power spectrum by approximately $1\%$, independently of scale, between $z=99$ and $z=49$. Previous work \citep[AHB13;][]{Brandbyge_2008, Viel_2010} avoided this discrepancy with linear theory by starting their simulations at a lower redshift ($z=49$), but this leads to inaccuracy in the CDM power spectrum \citep{Heitmann:2010}. This problem is more severe for lower neutrino masses, which require a lower starting redshift to avoid shot noise. To demonstrate explicitly that it is early-time shot noise which leads to this discrepancy, we performed a hybrid simulation where all neutrinos are followed by the particle component (\texttt{HYBRID-v5000} in Table \ref{tab:simulations}). This simulation is identical to the linear response simulation before the particle switch-on time of $z=1$ and evolves as a particle simulation thereafter. Like the simulations shown in Figure \ref{fig:matter_power}, it produces a matter power spectrum in good agreement with CAMB. We have checked that re-running the \texttt{PARTICLE} simulation with the tree force disabled, which mitigates discreteness noise by smoothing the gravitational force on scales of the PM grid, reproduces the correct large-scale power. Finally, we remind the reader that we have rebuilt the gravitational tree every timestep; simulations which do not do this may instead observe an increase in power beyond the expected linear theory.

Figure \ref{fig:minimal_mass} shows the matter power spectra from simulations with $M_\nu = 0.06$~eV $z=0$ and $z=0.5$. These low neutrino masses will become increasingly important as the upper limit on $M_\nu$ is reduced. We show linear response and particle simulations, but omit a hybrid simulation as for this neutrino mass $ < 4\%$ of the neutrinos have an initial velocity lower than $850$ km/s. Assuming a normal hierarchy, the three mass states are $0.05, 0.009, 0.001$ eV, respectively. The particle simulation has the same background evolution, but includes perturbations from only one neutrino mass state with $0.05$ eV. Neutrinos with a mass of $0.009$ eV have a free-streaming length of $860$ h/Mpc at $z=0$, larger than our box size, so this is a good approximation. Curiously, Figure~\ref{fig:minimal_mass} shows that at $z=0$ our simulations lack the usual upturn in the relative power spectrum at $k > 3$ h/Mpc. This is a chance feature of our particular realisation; we have confirmed that a different structure seed produces the usual spoon shape at all redshifts.

The linear response and particle simulations both agree with linear theory for $k < 0.1$ h/Mpc, further demonstrating that our simulations are able to reproduce the results of linear theory on large scales even for the lowest neutrino mass allowed by oscillation experiments. The particle simulation and the linear response simulation are discrepant at the level of $0.2\%$, a small difference compared to both observational and theoretical errors in structure formation.


\subsection{Halo Mass Functions}
\label{sec:halomass}

\begin{figure}
  \includegraphics[width=0.45\textwidth]{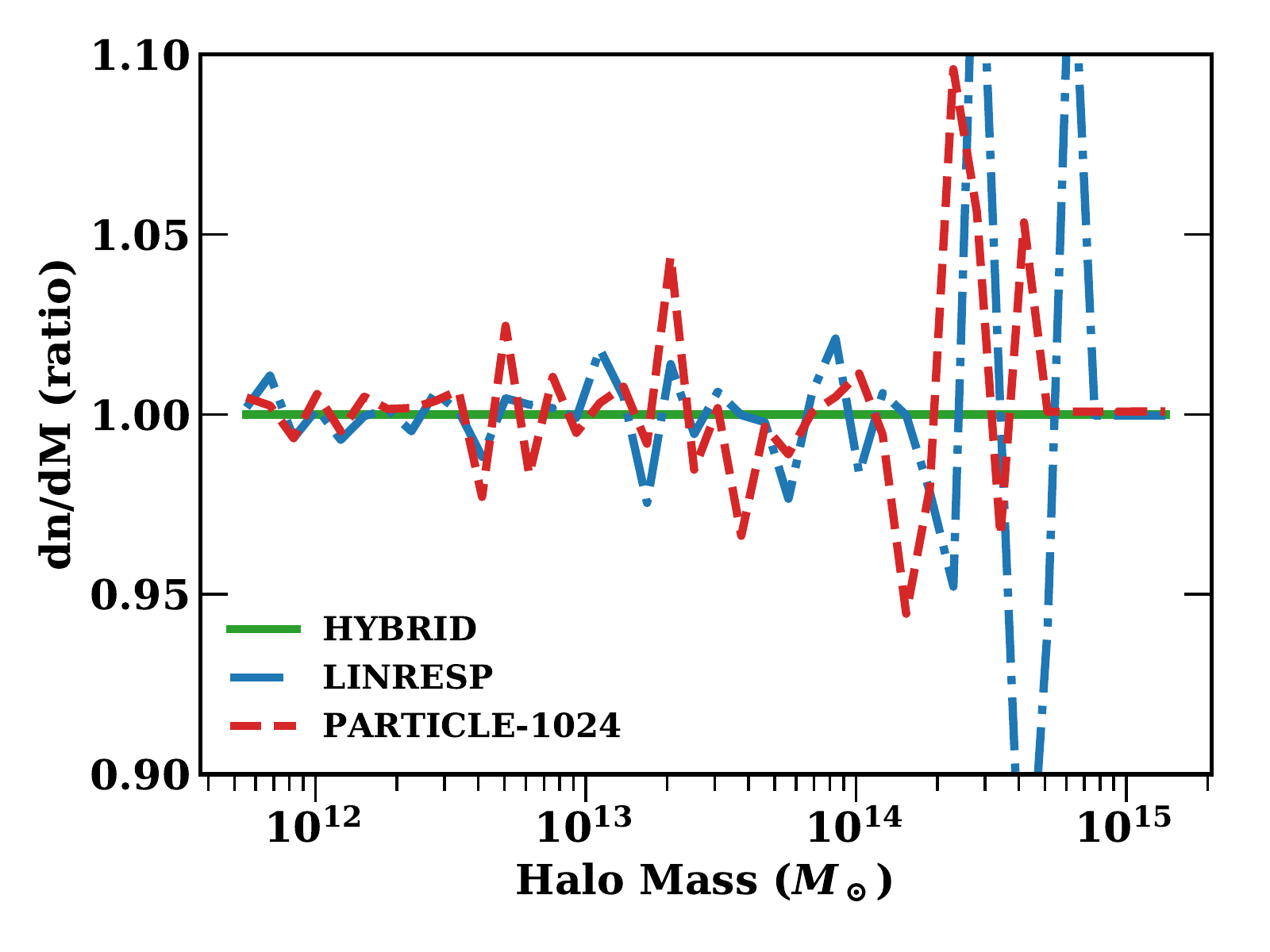}
\caption{Halo mass functions at $z=0$ for our \texttt{HYBRID}, \texttt{PARTICLE-1024} and \texttt{LINRESP} simulations, divided by the halo mass function from the \texttt{HYBRID} simulation. $M_\nu = 0.4$ eV.}
  \label{fig:halomass}
\end{figure}

Figure~\ref{fig:halomass} shows the halo mass functions for our particle, linear response and hybrid simulations. In order to easily compare the different simulations, halos are composed only of CDM particles; neutrino particles are excluded. Our box is too small to produce a statistical sample of large halos or compare the halo mass functions to theoretical expectations, so we show only the ratio to the halo mass function in the \texttt{HYBRID} simulation.

The simulated mass functions are nearly the same in each neutrino simulation method.
There is a few percent scatter between individual mass bins, especially at larger masses where the total number of halos in the box is small. All our methods of simulating neutrinos thus produce similar results for strongly non-linear structures.

\subsection{Clustering of neutrinos}
\label{sec:nupower}

\subsubsection{Correlation of fast neutrinos with the matter field}

\begin{figure}
\includegraphics[width=0.45\textwidth]{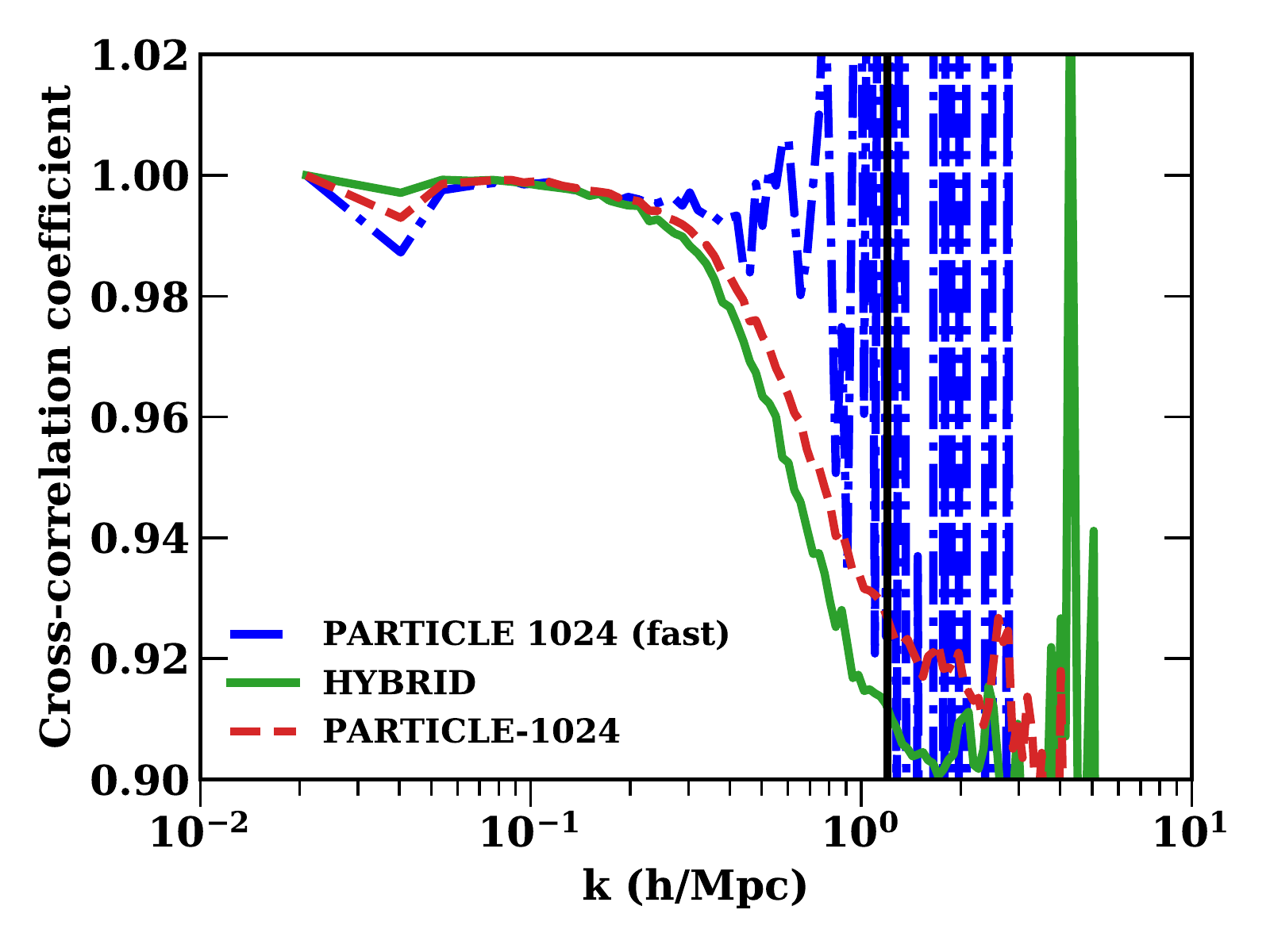}
  \caption{The cross-correlation coefficient between neutrinos and dark matter using the particle and hybrid simulation methods. Shot noise has been subtracted from the neutrino power spectrum. For the hybrid simulation, only the correlation between the cold dark matter and slow neutrino particles is shown. For the particle simulation, we have split the neutrino particles by their initial momentum before computing their cross-correlation with CDM. Thus the dot-dashed PARTICLE 1024 (fast) line shows only the neutrino particles which would be followed with the LRA in a hybrid simulation, while the dashed PARTICLE 1024 line shows all neutrino particles in that simulation.
  The vertical black line shows the scale where shot noise dominates the particle simulation.
  }
  \label{fig:cross-corr}
\end{figure}

As well as the LRA, we make the additional approximation \eqref{eq:phases} in the linear response integral, so that we need not keep track of the history of the full 3-dimensional matter density field. This approximation implies that fast neutrinos are completely correlated with the dark matter. It is certainly accurate on linear scales, where modes are decoupled. We can further check its consistency with a particle neutrino simulation. Figure~\ref{fig:cross-corr} shows the cross-correlation coefficient between neutrino particles and CDM, computed with NBodyKit, in the \texttt{HYBRID} and \texttt{PARTICLE-1024} simulations, demonstrating that it is indeed close to unity. We have subtracted shot noise from the neutrino power spectrum, so that the cross-correlation coefficient is defined to be

\begin{equation}
 r = \frac{\left<\delta_\mathrm{CDM} \delta_\nu \right>}{\sqrt{\left<\delta_\mathrm{CDM} \delta_\mathrm{CDM} \right>\left(\left<\delta_\nu \delta_\nu \right> - P_\mathrm{shot}\right)}}
\end{equation}
where $P_\mathrm{shot} = 300^3 / N_\mathrm{part}^3$ is the shot noise from the neutrino particles. On small scales where $P_\nu \ll P_\mathrm{shot}$ the power spectrum is hard to measure accurately, leading to numerical noise.

We further show the cross-correlation between the CDM and the ``fast'' and ``slow'' neutrino particles. The fast neutrino particles are correlated with the CDM at better than $99\%$ on all scales where shot noise permits the power spectrum to be reliably computed, with the exception of a single bin at $k \approx 0.035$ $h$/Mpc. In a hybrid simulation these particles would be followed using the LRA. The high cross-correlation between the particles and the CDM thus justifies Eq.~\eqref{eq:phases} for hybrid simulations. Even initially slow neutrino particles from a hybrid simulation are correlated at better than $99\%$ for $k < 0.3$ $h$/Mpc, which extends to the mildly non-linear regime. Smaller scales are still correlated at the $90\%$ level.

\subsubsection{Neutrino power spectrum}

\begin{figure*}
\includegraphics[width=0.45\textwidth]{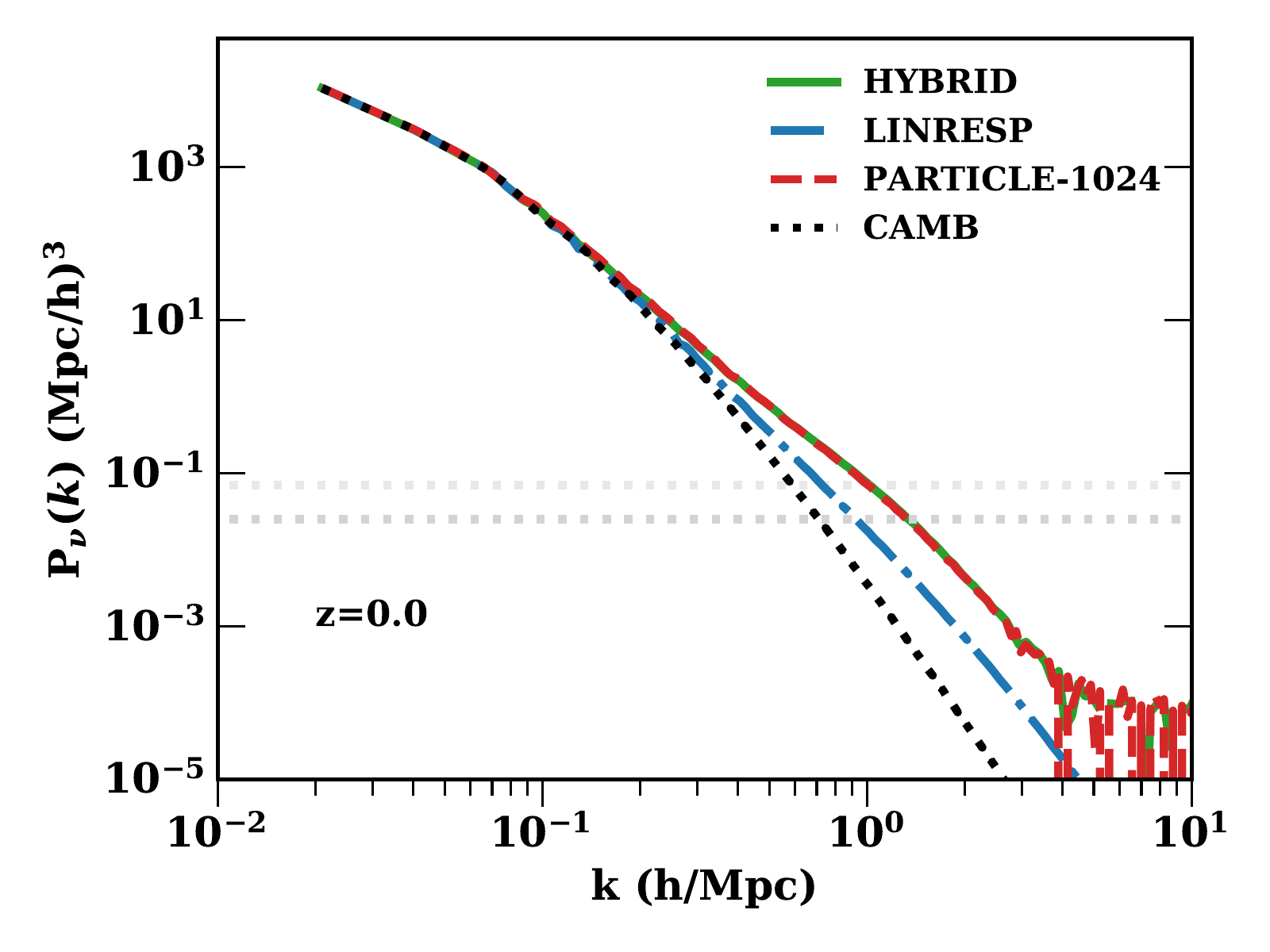}
\includegraphics[width=0.45\textwidth]{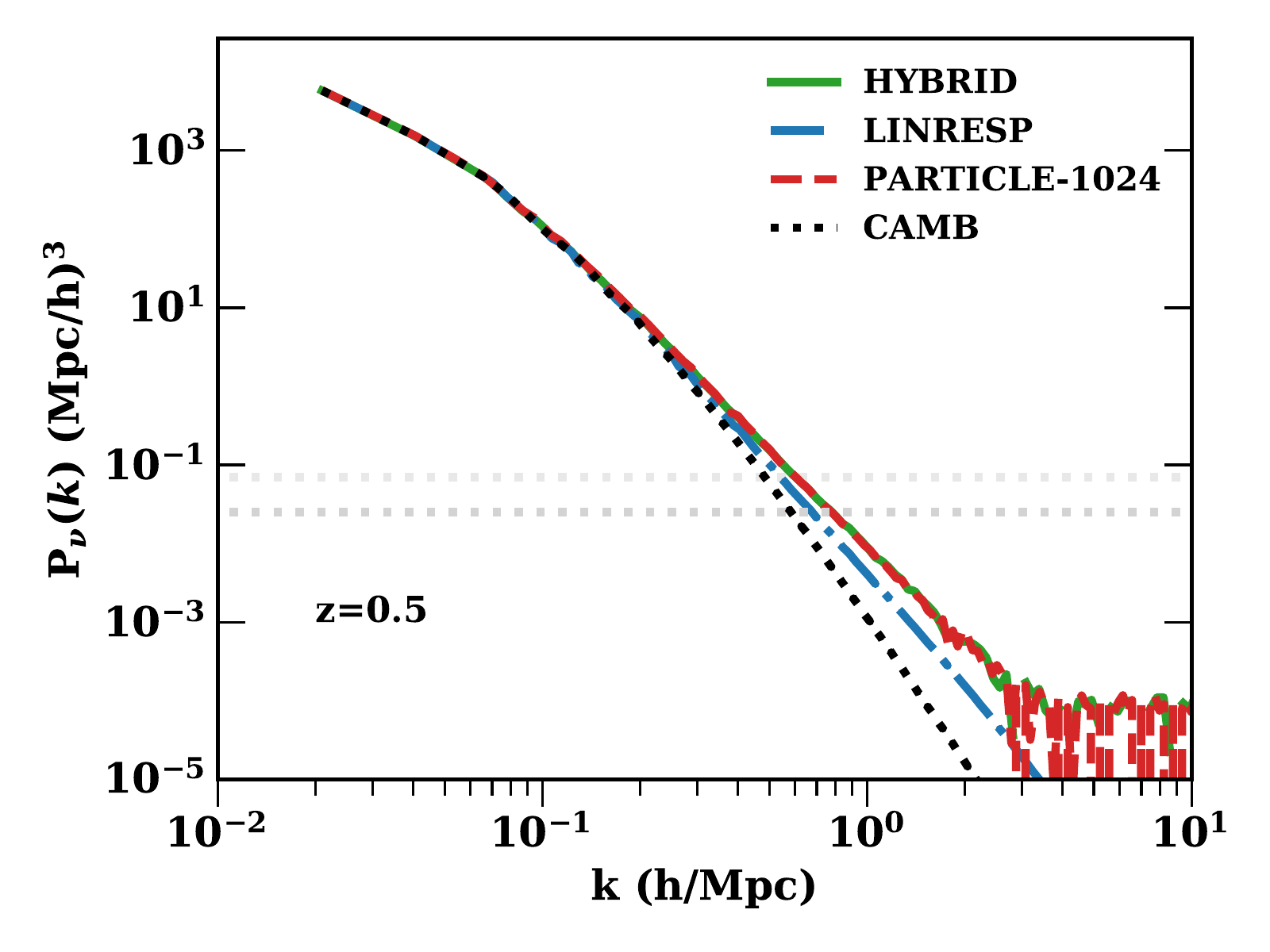}
  \caption{The neutrino power spectrum with massive neutrinos ($M_\nu = 0.4$ eV) for simulations using linear response (\texttt{LINRESP}) hybrid (\texttt{HYBRID}) and particle (\texttt{PARTICLE-1024}) methods. (Left) At $z=0$. (Right) At $z=0.5$. We have subtracted shot noise from the particle and hybrid simulations. The heavier dashed grey curve shows the level of shot noise subtracted from the particle simulation, while the lighter dashed grey curve shows the level subtracted from the hybrid simulation. We show the linear theory neutrino power spectrum from CAMB for comparison. There is good agreement between the hybrid and particle simulation methods.}
  \label{fig:neutrino_power}
\end{figure*}

\begin{figure}
\includegraphics[width=0.45\textwidth]{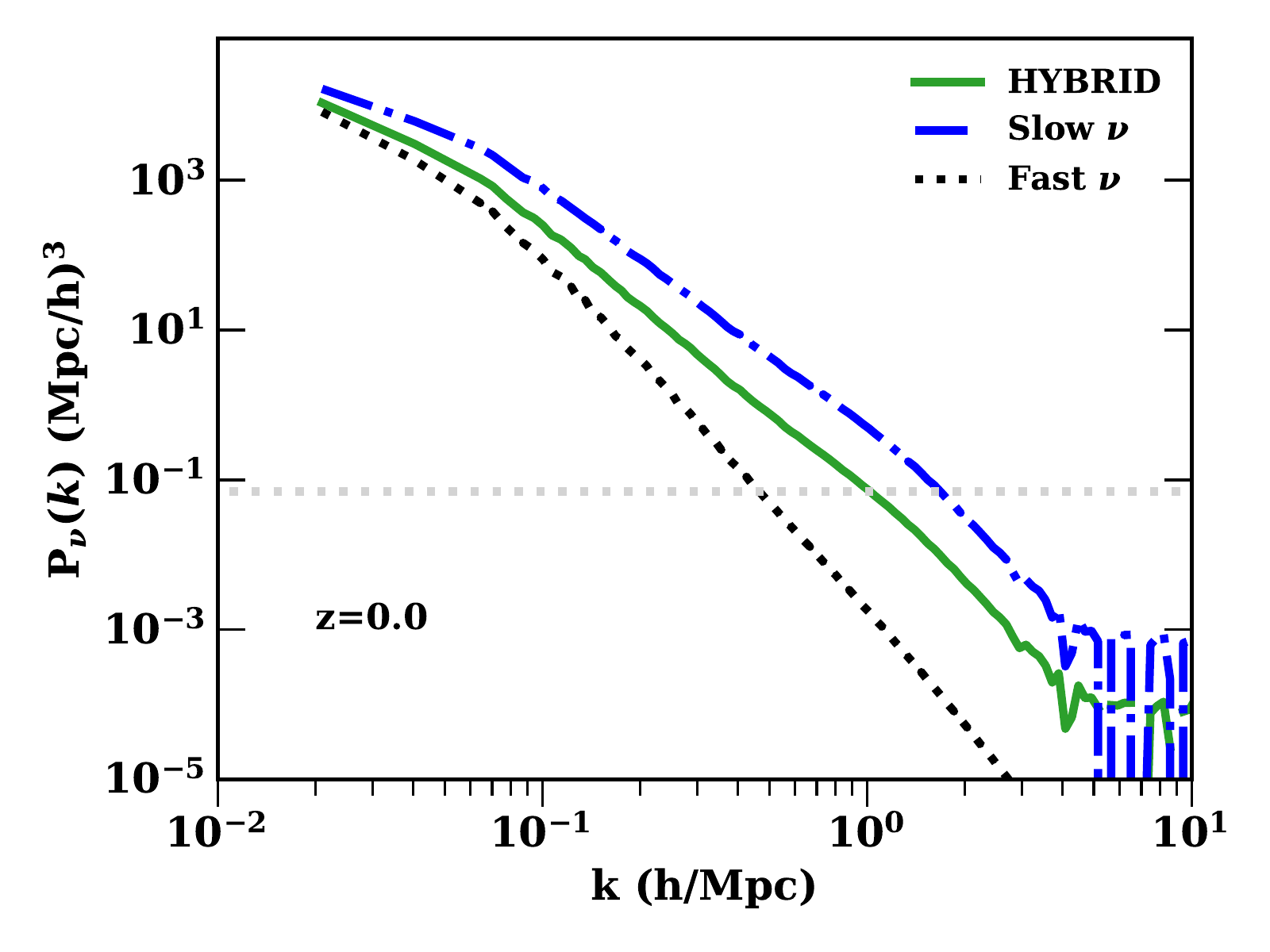}
  \caption{The neutrino power spectrum for the hybrid simulation (\texttt{HYBRID}) at $z=0$, split into fast (analytic) and slow (particle) neutrino components. Shot noise has been subtracted at the level shown by the grey line.}
  \label{fig:neutrino_power_split}
\end{figure}

\begin{figure}
  \includegraphics[width=0.45\textwidth]{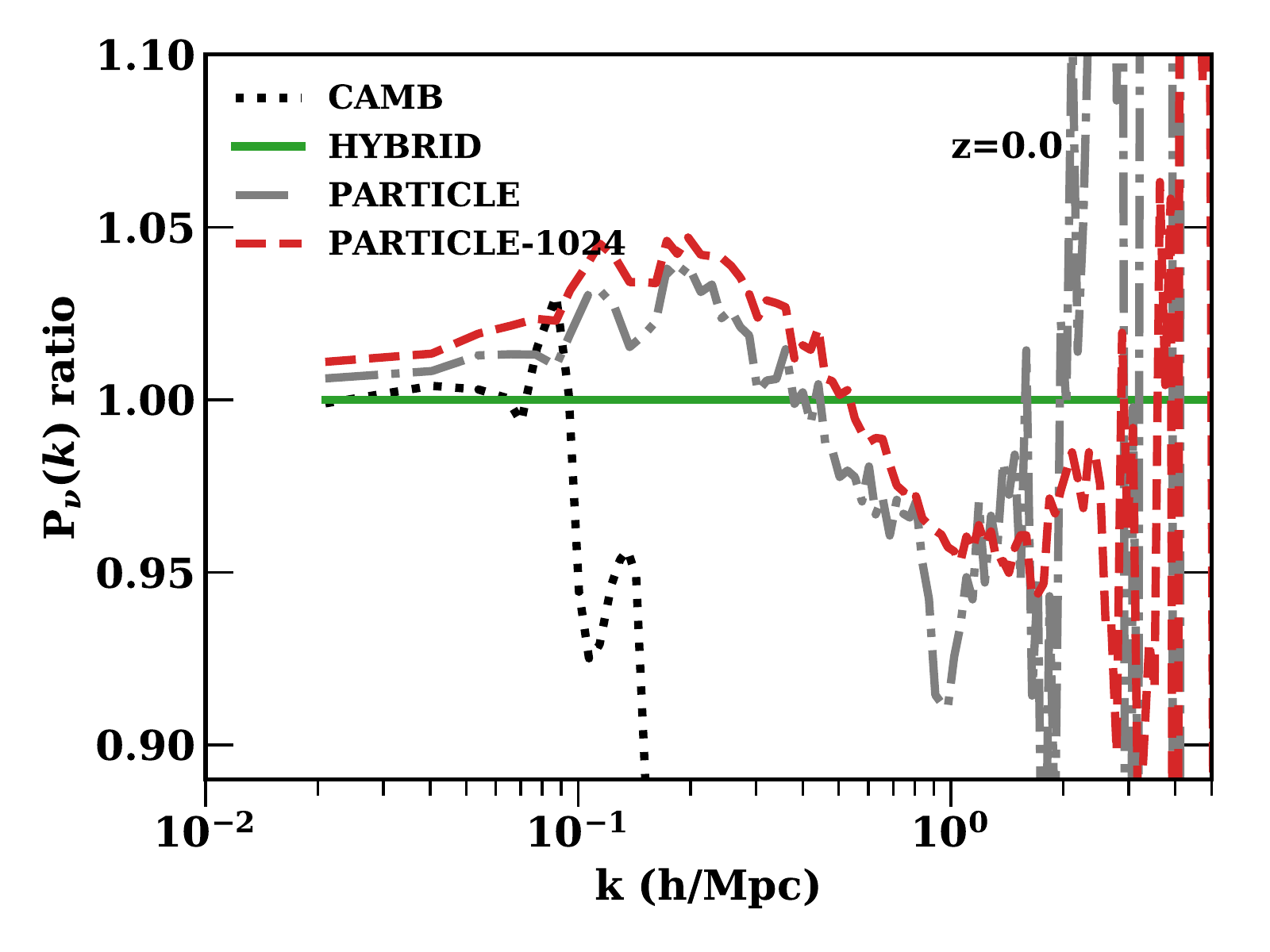}
    \caption{Ratios of neutrino power spectra to the results of the default hybrid simulation. Shown are the default (\texttt{HYBRID}), neutrino particle simulations with $512^3$ neutrino particles (\texttt{PARTICLE}) and $1024^3$ neutrino particles (\texttt{PARTICLE-1024}). Power spectra have been smoothed with an 4-pt rolling average to reduce scatter and shot noise has been subtracted. We also show the ratio of \texttt{HYBRID} to linear theory from CAMB, which agrees at the percent level on large scales.}
  \label{fig:hybparticle}
\end{figure}

\begin{figure}
  \includegraphics[width=0.45\textwidth]{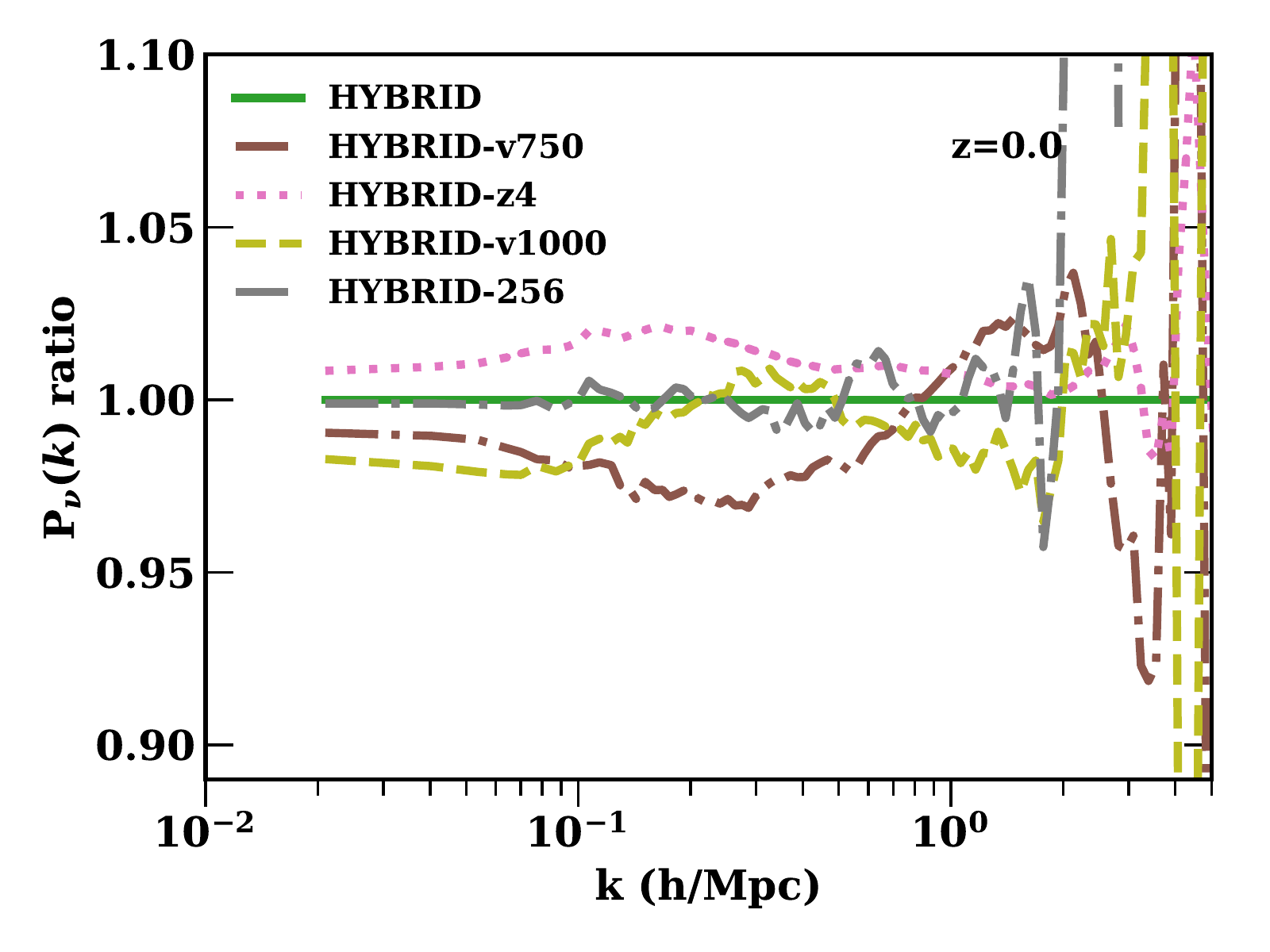}
  \caption{Neutrino power spectra from various hybrid simulations compared to the results of the default hybrid simulation (\texttt{HYBRID}). \texttt{HYBRID} has a cut-off velocity of $850$ km/s, a neutrino switch-on time of $z=1$ and $512^3$ neutrino particles. Other simulations shown have neutrino cut-off velocities of $1000$ km/s \texttt{HYBRID-v1000} and $750$ km/s (\texttt{HYBRID-v750}), a switch-on time of $z=4$  (\texttt{HYBRID-z4}) and $256^3$ neutrino particles (\texttt{HYBRID-256}). The ratio has been smoothed with an 4-pt rolling average to reduce mode-to-mode scatter and shot noise has been subtracted.}
  \label{fig:vcrit}
\end{figure}

Figure~\ref{fig:neutrino_power} shows the neutrino power spectrum at $z=0$ and $z=0.5$, demonstrating each of the neutrino simulation methods, as well as the linear theory power spectrum from CAMB. For the hybrid simulation we have computed the total neutrino power spectrum (assuming both neutrino components are completely correlated) as the weighted sum of the power spectrum of the fast and slow components $P^{1/2}_\nu = f_\mathrm{fast} P^{1/2}_\mathrm{fast} + f_\mathrm{slow} P^{1/2}_\mathrm{slow}$, where $f_\mathrm{fast} + f_\mathrm{slow} = 1$. Figure~\ref{fig:cross-corr} shows that this is a good approximation for $k < 0.5$ h/Mpc. For $k > 0.5$ h/Mpc, the slow neutrinos contribute $> 95\%$ of the total power, so that a $10\%$ decorrelation would over-estimate the neutrino power by only $0.5\%$. Figure~\ref{fig:neutrino_power_split} shows the neutrino power spectra for each individual component, showing the increased clustering for the slow neutrinos.

Neutrino particle shot noise has been subtracted from both the particle and hybrid simulation. The shot noise for the hybrid simulation is $P_\mathrm{shot} = 0.346^2\times (300 /512)^3 $ (Mpc/$h)^3$, where the factor of $0.346$ is due to the reduced matter density in particle neutrinos. The particle simulation has $P_\mathrm{shot} = (300 /1024)^3$ (Mpc/$h)^3$. In both simulations, the neutrino power is recoverable even two orders of magnitude below the shot noise level, indicating that there is little structure formation arising purely from neutrino shot noise.


At $z \geq 1$ all three methods are in good agreement, consistent with the results of AHB13. At $z = 0$ and $z=0.5$, the hybrid neutrino simulation and neutrino particle simulation are in good agreement. However, the linear response method shows reduced power on small scales. This is in agreement with the expectations from Section~\ref{sec:hybrid} and demonstrates that our hybrid method indeed resolves the discrepancy between the linear response method and the particle method. We show a comparison to the linear theory prediction from CAMB. For $k < 0.1$ $h$/Mpc, both the HYBRID and LINRESP (not shown) simulations agree with CAMB at the $1\%$ level. On smaller scales, non-linear growth in the CDM leads to more power than the linear prediction.

Figure~\ref{fig:hybparticle} compares the neutrino power spectrum from the particle and hybrid methods in more detail. There are two interesting features: a $5\%$ deficit of power in the particle simulations at $k \approx 1$ $h$/Mpc, and a $5\%$ increase in power at $k \approx 0.2$ $h$/Mpc. The origins of these features are unclear. However, we consider changes of $5\%$ in the neutrino power spectrum an acceptable degree of convergence; $f_\nu = 0.02$ for $M_\nu = 0.3$ eV and $P_\nu \ll 0.01 P_\mathrm{M}$ at $k = 0.1$ h/Mpc, so there is a maximal effect on $P_\mathrm{M}(k)$ of $10^{-5}$.



To evaluate the possibility that the differences between the particle and hybrid methods are due to non-linear growth in neutrinos at $z > 1$, Figure \ref{fig:vcrit} shows the \texttt{HYBRID-z4} simulation. \texttt{HYBRID-z4} changed the initial neutrino switch-on time from $z=1$ to $z=4$. There is a moderate discrepancy, with a $z=4$ switch-on time leading to $1-2\%$ more power than a $z=1$ switch-on time. This suggests that non-linear growth at $z> 1$ could account for up to half of the increased neutrino power in the particle simulation. We caution though, that non-linear growth between $z=1$ and $z=4$ could also be unphysical, as there are still clear shot noise residuals in \texttt{HYBRID-z4} at $z = 2$.

Figure~\ref{fig:vcrit} shows how the neutrino power spectrum from hybrid simulations depends on the model parameters. We show simulation outputs at $z=0$, but have checked that higher redshifts produce similar results. In particular, we have varied the neutrino switch-on time, the critical neutrino velocity and the number of neutrino particles (and thus the level of shot noise). The total matter power spectra for all these simulations agreed to $0.05\%$ for $k < 3$ h/Mpc. Increasing the critical neutrino velocity to $v_\mathrm{crit} = 1000$ km/s (\texttt{HYBRID-v1000}) from $850$ km/s did not alter the neutrino power spectrum by more than $1\%$ for $k > 0.08$ $h$/Mpc, as all non-linear growth occurs in neutrinos with unperturbed velocities less than $850$ km/s. By contrast, when the critical velocity is reduced to $750$ km/s (\texttt{HYBRID-v750}), the neutrino power spectrum is reduced by $3\%$, indicating moderate non-linear growth in these velocity bins. These results confirm the calculations of Section~\ref{sec:validity}.

Finally, we show \texttt{HYBRID-256}, a hybrid simulation where the neutrino particle load has been decreased by a factor of $8$, to $256^3$. This changed the neutrino power spectrum by less than $1\%$ and shows that we are well converged with respect to mass resolution and neutrino shot noise at $z < 1$.
The effects of changing the hybrid simulation parameters are not large, and our simulations appear well converged.

\section{Conclusions}
\label{sec:conclusion}

We have extended the linear response neutrino simulation method from \cite{AHB} to better account for non-linear growth in the neutrino component and thus reproduce the non-linear neutrino power spectrum as well as the non-linear matter power spectrum.
Our improved method is a hybrid: initially fast-moving neutrinos are followed as before using a linear response method, while initially slow-moving neutrinos, which can be captured by CDM halos, are followed using particles at late times. Neutrinos are followed analytically at early times, allowing the hybrid method to avoid the impact of shot noise with a much lower particle load. We show that our new hybrid method reproduces the non-linear matter power of the linear response simulations, as expected from \cite{AHB}, while also reproducing the significantly large neutrino power spectrum seen in a converged particle simulation at $z=0$. We show that the hybrid method agrees well with CAMB when structure growth is linear. Since only a fraction of the neutrino matter density is followed by neutrinos, we show that converged results can be obtained with a relatively small neutrino particle load. Our simulation code is publicly available, both integrated into the simulation code MP-\gadget and as a series of patches to \gadget-2 \footnote{\url{https://github.com/sbird/kspace-neutrinos}}. We showed that a neutrino switch-on redshift $z=1$ and a critical neutrino velocity of $850$ km/s include the majority of non-linear neutrino velocity shells, as well as showing good convergence properties for $M_\nu \leq 0.4$ eV.

Most simulators wishing to compare to observational surveys can use the linear response method, as it is computationally efficient and still reproduces well the properties of the total matter density, which are the directly observable quantities. It is, for example, completely sufficient for computing lensing convergence power spectra \citep{McCarthy_2018, Liu_2017}.

For simulators wishing instead to investigate the structure of the neutrino component, our hybrid method provides much improved accuracy. Simulations using our hybrid method could be used to investigate, for example, the distribution of neutrino matter around collapsed objects \citep{FVN_2013}, neutrino wakes \citep{Inman_2015}, the neutrino bispectrum \citep{Furhrer_2015, Ruggeri_18} or the distribution of massive neutrinos in cosmic voids \citep{Banerjee_2016}. Our code will also be useful to run large-box simulations to investigate halo bias in massive neutrino cosmologies \citep{Loverde_14, Chiang_17}. A further extension of our method could be to suppress shot noise even further by using the initial conditions method of \cite{Banerjee_2018}.

Our linear response simulations have computational costs similar to pure cold dark matter simulations. Our hybrid simulations required about twice the computational time of a linear response simulation, for a neutrino particle load equal to that of the CDM. This compares well to the extra cost of the particle simulations, which was a factor of $10$ in CPU time and $8$ in memory for a fully converged simulation. Simulators may consider further reducing the computation cost of the hybrid simulations by reducing the neutrino particle load below that of the CDM. We found that in our simulations a particle load of $256^3$, $1/8$th that of CDM, gave identical results for the neutrino power spectrum. Our hybrid linear response-particle simulations are thus substantially faster than pure particle simulations, allowing simulation suites to sample a cosmological parameter space with massive neutrinos substantially more finely. Furthermore, the decreased streaming velocity of neutrino particles may make (at least approximate) ``zoom-in'' simulations of individual halos possible, as long as the halos are sufficiently small that any neutrinos slow enough to be captured can be included in a small box.

Finally, we note that cosmological surveys have now reached a level of sensitivity where even the minimal neutrino mass can substantially alter derived parameters \citep{Calabrese_2017}, and thus the inclusion of massive neutrinos (and radiation in the background) should become standard for all simulators, including those building cosmological emulators \citep{Lawrence_2017}. For these mass ranges including the neutrino mass hierarchy is important, and removing all neutrino particle shot noise prohibitively expensive. Both problems are avoided by our linear response or hybrid methods.

\section*{Acknowledgements}

We thank Arka Banerjee for sharing simulation data. We thank Jeremy Tinker and Derek Inman for useful discussions.
This research project was conducted using computational resources
at the Maryland Advanced Research Computing Center (MARCC). SB was supported by NASA through
Einstein Postdoctoral Fellowship Award Number PF5-160133. JL is supported by an NSF Astronomy and Astrophysics Postdoctoral Fellowship under award AST-1602663. This work used the Extreme Science and Engineering Discovery Environment (XSEDE), which is supported by NSF grant ACI-1053575.
\appendix

\section{Manual}
\label{sec:manual}

In this Appendix, we briefly describe the parameters of the linear response neutrino method. A similar description may be found in the README of the code repository: \url{https://github.com/sbird/kspace-neutrinos/}\,. Our neutrino integrator has been altered to be a stand-alone module, largely independent of the underlying N-body code. To aid integration, we have included copious comments and unit tests. A script is provided in the repository which downloads and patches a fresh copy of \gadget-2 to include massive neutrinos: the ``apply-patches'' script in the \gadget-2 subdirectory.

Table \ref{tab:parameters} shows a list of the required parameters, as well as brief descriptions. The number of extra parameters required is small. Three parameters are required to specify the initial power of the neutrino component, using a CAMB or CLASS transfer function file. Three parameters are required to specify the masses of the three active neutrino species.
There is a global switch enabling the hybrid neutrino model. Note that the matter power spectrum is extremely well converged by the linear response method alone. The hybrid neutrino model includes two additional parameters: the critical velocity below which neutrinos are particles, and the neutrino switch-on time, after which neutrinos are actively gravitating. The default values of these parameters are justified in Section~\ref{sec:hybrid}, and may be altered as desired for the problem of interest.
Note that the critical velocity used in MP-\gadget~should match that set in the initial conditions code. In this work we also used a neutrino particle load $8$ times smaller than the CDM particle load, which was sufficient to produce a converged neutrino power spectrum on the scales of interest.

\begin{table*}
\begin{center}
\begin{tabular}{|l|l|}
\hline
    Parameter & Description \\
\hline
KspaceTransferFunction   & CMB transfer functions, used to compute the neutrino integration. \\
TimeTransfer             & Scale factor of the CMB transfer functions. \\
InputSpectrumUnitLengthincm   & Units of the CAMB transfer function in cm. \\
MNue, MNum, MNut &  Three neutrino masses. The measured mass splittings are not enforced. \\
Vcrit            & Critical velocity below which the neutrinos are particles, if hybrid neutrinos are on. \\
NuPartTime       & Scale factor at which the particle neutrinos start to gravitate, if hybrid neutrinos are on. \\
HybridNeutrinosOn       & Switch to enable hybrid neutrinos. \\
\hline
\end{tabular}
\end{center}
\caption{Table of code parameters, with brief descriptions.}
\label{tab:parameters}
\end{table*}

As documented in \cite{Springel_2005} and the \gadget-2 manual, \gadget-2 and some versions of \gadget-3 output snapshots mid-timestep. This is implemented by drifting all particles (even those not currently active) to the desired output time. However, particles are not kicked to update their momenta, so that the output particle velocities are those from the last active timestep. For neutrino particles, whose clustering is intimately tied to their total momentum, restarting from a snapshot using \gadget-2 or later will introduce an error in the $z=0$ power spectrum. Users should restart their simulations, if necessary, from restart files. This does not apply for MP-\gadget, which we have modified so that snapshots always occur at the end of a PM timestep, when all particles are active. Finally, when using \gadget-3 and neutrino particles, the force tree should be rebuilt every timestep by setting the ``TreeDomainUpdateFrequency'' parameter to zero to avoid the inaccuracy described in Section~\ref{sec:partnuimprovements}.

\section{Initial Conditions}
\label{sec:initcond}


In this Appendix, we detail improvements to the accuracy of our simulation initial conditions since AHB13.
Following Lagrangian perturbation theory \citep{Zeldovich_1970, Scoccimarro_1998},
the particle velocities and displacements are related:
\begin{equation}
v(k) = a H(a) \frac{d \log D(a)}{d \log a} \delta(k)\,.
\label{eq:vel_prefac}
\end{equation}
$D(a)$ is the linear growth and $H = \dot{a}/a$ the Hubble function.

In AHB13, the Hubble function used in Eq.~\eqref{eq:vel_prefac}
neglected radiation density. Furthermore, following \cite{Bouchet:1995}, we
approximated the derivative of the linear growth function, $\frac{d \log D(a)}{d \log a}$, by
\begin{equation}
\frac{d \log D(a)}{d \log a} \approx \left(\frac{\Omega_M a^{-3}}{\Omega_M  a^{-3} + \Omega_\Lambda}\right)^{0.6} \approx 1\,.
\end{equation}

This approximation is valid in matter domination, but again neglects the radiation density,
which becomes non-negligible at $z > 50$. Both of these approximations are especially notable
when simulating massive neutrinos, because at high redshift neutrinos are slightly relativistic,
and thus the background density depends slightly on the neutrino mass. In practice the error
induced by each approximation partially cancels, leaving an under-estimation of the effect of
massive neutrinos on structure formation by $\sim 2 \%$, visible in, for example,
Figure 4 of AHB13\footnote{We thank Francisco Villaescusa-Navarro for first pointing these problems out to us.}.

In the simulations presented here we use both the full Hubble function
and obtain $\frac{d \log D(a)}{d \log a}$ by numerically solving
the linear growth equation \citep{Peebles:1993}:
\begin{equation}
\frac{d}{da}\left(a^3 H(a) \frac{d D(a)}{da}\right) - \frac{3}{2} \frac{H_0^2\,D(a)}{a^2 H(a)} \left(\Omega_\mathrm{CDM} + \Omega_\mathrm{b}\right)= 0\,.
\label{eq:lineargrowth}
\end{equation}
The initial conditions for this differential equation are equal to the growing mode (and derivative) in a matter-radiation universe, set at $z \gg 100$ \citep{Groth:1975}:
\begin{equation}
  D(a_i) = \Omega_\gamma + \Omega_\nu + \frac{3}{2} \left(\Omega_\mathrm{CDM} + \Omega_\mathrm{b}\right) a_i\,.
\end{equation}
The above is appropriate for our simulations, because our simulation box size is smaller than the neutrino free-streaming scale at our initial redshift, and thus neutrinos do not cluster. For simulations which probe large scales the scale-dependent growth rate should be used, as in \cite{OLeary_2012, Zennaro_2017}. We have checked explicitly that for our box size the scale-dependent and scale-independent growth rates are the same.

Our initial conditions are generated using the CAMB CDM + baryon transfer function at $z=99$. An alternative is to generate initial conditions using the $z=0$ transfer function, scaled to the initial redshift by the (sub-horizon) linear growth function, $D(z_\mathrm{ic})$. This can be used to account for background radiation density, if radiation is not included in the background evolution, or for radiation perturbations and other relativistic effects on the scale of the horizon at $z=99$. For massive neutrinos the rescaling is scale-dependent and requires computing equations for a two-fluid system, as neutrinos are no longer free-streaming on the scales of interest at $z=0$ \citep{Zennaro_2017}.

For our simulations the effect of radiation on the background expansion is included and the simulation box is less than the horizon size. On sub-horizon scales, the $z=0$ matter power spectrum rescaled to $z=99$ correctly accounting for neutrinos is thus identical to the $z=99$ matter power spectrum we use for initialization, making the rescaling procedure trivial. Any non-trivial rescaling would introduce an error in the $z > 0$ power spectra due to the mismatched growth functions, which increases at higher redshift. We found that for a massless neutrino simulation which does not include radiation in the background evolution this error is $0.5\%$ on linear scales at $z=9$ and moderately larger on non-linear scales. This (small) error is easy to avoid by using the correct background evolution, as we have done in this paper.

\label{lastpage}

\bibliography{neutrinos}

\end{document}